\documentclass[aps,pra,twocolumn,a4paper,showpacs,superscriptaddress,floatfix,10pt]{revtex4}
\usepackage[pdftex]{graphicx,color}% Include figure files
\usepackage{dcolumn}% Align table columns on decimal point
\usepackage{bm}% bold math

\usepackage{amsfonts}
\usepackage{amssymb}
\usepackage{amsmath}
\usepackage{amsthm}

\usepackage{subfigure}
\usepackage{rotate}

\usepackage{pdfsync}

\usepackage[pdftex]{hyperref}

\let\oldmarginpar\marginpar
\renewcommand\marginpar[1]{\-\oldmarginpar[\raggedleft\tiny #1]%
{\raggedright\tiny #1}}

\DeclareMathOperator{\Tr}{Tr}

\newcommand{\bra}[1]{\langle#1|}
\newcommand{\ket}[1]{|#1\rangle}
\newcommand{\braket}[2]{\langle#1|#2\rangle}

\graphicspath{{figs/}}

\begin{document}

\title{Thermal inclusions: how one spin can destroy a many-body localized phase}

\author{Pedro Ponte}
\affiliation{Perimeter Institute for Theoretical Physics, Waterloo, Ontario N2L 2Y5, Canada}
\affiliation{Department of Physics and Astronomy, University of Waterloo, Ontario, N2L 3G1, Canada}

\author{C. R. Laumann}
\affiliation{Department of Physics, Boston University, Boston, MA 02215, USA}

\author{David A. Huse}
\affiliation{Department of Physics, Princeton University, Princeton, NJ 08544, USA}

\author{A. Chandran}
\email{anushyac@bu.edu}
\affiliation{Department of Physics, Boston University, Boston, MA 02215, USA}

\date{\today}

\begin{abstract}

Many-body localized (MBL) systems lie outside the framework of statistical mechanics, as they fail to equilibrate under their own quantum dynamics.
Even basic features of MBL systems such as their stability to thermal inclusions and the nature of the dynamical transition to thermalizing behavior remain poorly understood.
We study a simple model to address these questions: a two level system interacting with strength $J$ with $N\gg 1$ localized bits subject to random fields.
On increasing $J$, the system transitions from a MBL to a delocalized phase on the \emph{vanishing} scale $J_c(N) \sim 1/N$, up to logarithmic corrections.
In the transition region, the single-site eigenstate entanglement entropies exhibit bi-modal distributions, so that localized bits are either ``on" (strongly entangled) or ``off" (weakly entangled) in eigenstates.
The clusters of ``on" bits vary significantly between eigenstates of the \emph{same} sample, which provides evidence for a heterogenous discontinuous transition out of the localized phase in single-site observables.
We obtain these results by perturbative mapping to bond percolation on the hypercube at small $J$ and by numerical exact diagonalization of the full many-body system.
Our results imply the MBL phase is unstable in systems with short-range interactions and quenched randomness in dimensions $d$ that are high but finite.

\end{abstract}

\maketitle

\section{Introduction}
\label{sec:introduction}

There has been considerable recent theoretical and experimental activity in discovering and understanding the dynamical phases of generic isolated quantum systems.
The many-body localized (MBL) phase has been of particular interest, as it is a dynamical phase in which an isolated many-body quantum system is not able to act as a reservoir and bring its subsystems to thermal equilibrium \cite{Anderson:1958ly,Gornyi:2005lq,Basko:2006aa,Oganesyan:2007aa,Pal:2010gs,Nandkishore:2015aa,Altman:2015aa}.
MBL phases exhibit a complex of intriguing properties absent in conventional thermalizing phases, including an extensive set of localized conserved quantities (``l-bits'') \cite{Serbyn:2013rt,Huse:2014ac,Imbrie2016,Ros:2015rw,chandran2015constructing,Monthus:2016aa,Rademaker:2016aa}, slow entanglement dynamics \cite{Znidaric:2008aa,Bardarson:2012kl,Bauer:2013rz,Serbyn:2013uq}, dynamically protected long-range orders in highly excited states that are forbidden in equilibrium \cite{Huse:2013aa,Pekker:2014aa,Chandran:2014aa,Kjall:2014aa, Bahri:2015aa} among other interesting features \cite{Bar-Lev:2015aa,Devakul:2015aa,Torres-Herrera:2015aa,Gopalakrishnan:2016aa,Luitz:2016aa,Znidaric:2016aa,Altland:2017aa}.
MBL phases (or at least very slow dynamics approximating MBL) have been experimentally observed in cold atoms \cite{Kondov:2015aa,Schreiber:2015aa,Choi:2016aa,Bordia:2017aa}, trapped ions \cite{Smith:2016aa} and in diamond with nitrogen-vacancy centers \cite{Kucsko:2016aa}, in various regimes (one and two dimensions, with long-range interactions, with quasi-periodic potentials, etc.) and there are many more exciting experiments in the pipeline.

\begin{figure}[htbp]
\includegraphics[width=\columnwidth]{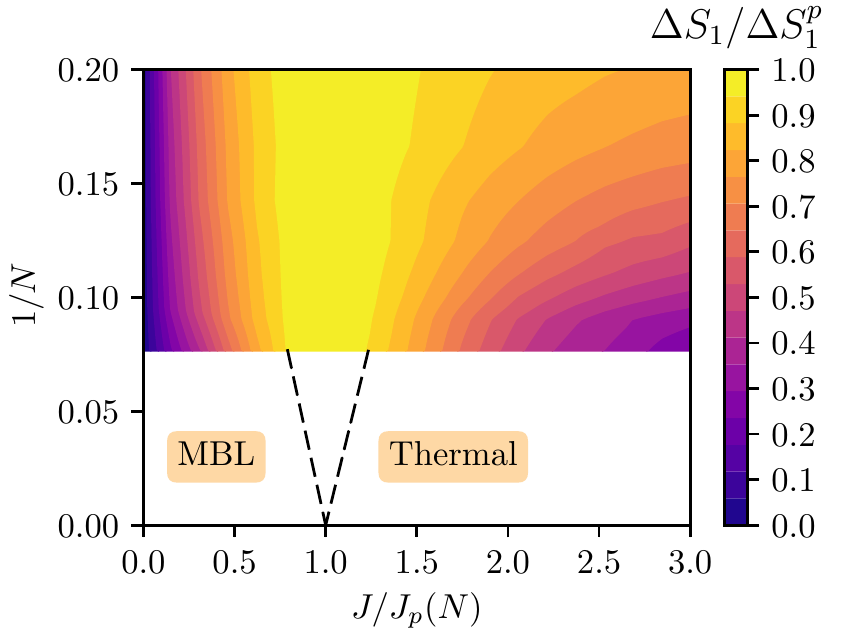}
\caption{
\label{fig:phase_diagram}
Schematic finite size phase diagram of the central spin model mapped out through the standard deviation $\Delta S_1$ over l-bits, eigenstates and samples of the single site entanglement entropy under the simplest assumption of a single transition at $J_c = J_p(N)$.
}
\end{figure}

While MBL phases have been essentially proven to exist in one dimensional systems \cite{Imbrie2016}, their existence and stability in higher dimensions remains controversial.
Ref.~\cite{Chandran:2016ac} studied the effect of a thermalizing boundary on a $d>1$-dimensional MBL phase and argued that the l-bits and other eigenstate measures of localization are unstable; instead, approximately conserved ``l$^*$-bits'' underlie what remains of localization in such systems.
Another study of the effect of thermal inclusions was presented in Ref.~\cite{De-Roeck:2017aa}, in which the authors argue that any nonzero density of large enough thermalizing inclusions will destroy the MBL phase in $d>1$, converting it to an extremely slow, but ultimately thermalizing regime.
The outline of their argument is as follows.
Consider a single thermal inclusion. If the inclusion is sufficiently coupled to the MBL phase in which it is embedded, then it will thermalize the localized degrees of freedom (l-bits) bordering the inclusion, and this allows the thermal inclusion to grow in size.
This then makes the inclusion a more effective ``bath" for l-bits that are further away.

In $d=1$, there is a critical value for the exponential decay length of the interaction between the inclusion and the l-bits beyond which this process runs away.
This was recently demonstrated numerically in Ref.~\cite{Luitz:2017aa}.
In $d>1$ systems with randomness, this process \emph{always} runs away so that there is a finite, but possibly extremely large, time-scale over which a putative MBL phase is thermalized by the nonzero density of rare thermal inclusions which arise for any strength of disorder.
This argument leaves open the possibility that MBL phases can remain stable in higher $d$ in nonrandom systems with quasiperiodic fields or potentials that do not produce such rare thermalizing regions.
And, as we address here, the MBL phase in a certain sense also regains its stability in the limit of infinite coordination number.

One of the motivations of this article is to study the key step in the destabilization process described above in $d>1$; that is, how a single inclusion interacting with the same strength $J$ with many l-bits on its border leads to their thermalization.
We show that a single two-level inclusion is sufficient to thermalize $N$ neighboring l-bits in the large-$N$ limit at any nonzero $J/W$, where $W$ is the strength of the random fields on the l-bits.
In order for such a ``central spin'' system to remain in the MBL phase in this large-$N$ limit, the interaction with the inclusion must be scaled down with increasing $N$.
Specifically, we find that the critical $J_{c}(N)$ separating the MBL phase for $J<J_{c}(N)$ from the thermalizing phase at $J>J_c(N)$ scales as $J_c(N) \sim W/N$ up to multiplicative $\log(N)$ corrections (see Fig.~\ref{fig:phase_diagram}).

At the transition for our central spin model, the interactions are crucial to the dynamics of the system but do not contribute to the system's $N\rightarrow\infty$ equilibrium thermodynamics.
Like other fully connected quantum spin glass models, the dynamical phase diagram simply does not coincide with the thermodynamic phase diagram \cite{Cugliandolo:2001aa,Laumann:2014aa,Burin:2016aa,Baldwin:2017aa,Mezard:1987aa}.
These models have been extensively studied in the context of quantum dots, ion traps and other long-range interacting spin systems; they provide a tractable analytic setting for studying many-body quantum dynamics.
In all of these models, there is a polynomially small in $N$ coupling strength $J_c(N)$ separating a MBL phase from the phases that fully or partially thermalize. At finite temperature, these models can also have glass phases where the system fails to fully thermalize due to divergent free energy barriers in the $N\rightarrow\infty$ limit.
It is important to make the distinction between arrested dynamics due to (1) {\it detuning}, which is the cause of Anderson and many-body localization, and due to (2) divergent free energy barriers.
In the latter case, the system can remain delocalized and function as a thermal bath for itself, only failing to thermalize a small number of collective degrees of freedom \cite{Baldwin:2017aa}.
Finally, for these fully connected models there are no destabilizing rare region effects, allowing for a stable MBL phase at $J < J_c(N)$.

In the sequel, we analyze the infinite temperature dynamical phase diagram of our central spin model using exact diagonalization with up to $N=13$ l-bits, and using small $J$ perturbation theory at large $N$.
Numerically, the single-site eigenstate entanglement entropy, energy level repulsion and many-body eigenstate participation ratios all support the existence of a MBL phase for $J < J_c(N) \sim W/N$ and a thermalizing (ETH) phase for $J > J_c(N)$ \footnote{We refer to the thermalizing phase as an ETH phase because it satisfies the eigenstate thermalization hypothesis \cite{Deutsch:1991ss,Srednicki:1994dw,Rigol:2008bh,DAlessio:2016aa,Borgonovi:2016aa}.}.
They also reveal many interesting features about the localized phase and the crossover region at $J \approx J_c(N)$.
In the localized phase, the mean single-site eigenstate entanglement entropy $[S_1]$ decreases as $1/N$, while the participation ratio distributions are $N$-independent.

In the crossover region, the l-bits are either ``on" (strongly entangled) or ``off" (weakly entangled) in eigenstates and the pattern of ``on" l-bits varies significantly between states of the same sample (and of course, between samples).
Thus, single-site observables are very heterogenous in real-space, in energy space and across disorder realizations in the crossover region, suggesting that they change discontinuously as $N\to \infty$, in line with recent proposals \cite{Yu:2016aa,Khemani:2016aa} that few-body observables are similarly discontinuous across the MBL transition in one dimension.

The perturbative analysis at small $J$ on the classical hypercube explains many of these numerical observations.
At second order, a typical initial configuration is resonant with $K \sim J^2N^2/W^2$ states in which two l-bits are flipped and the central spin is flipped.
These states are in turn resonant with $\sim K$ \emph{completely new} states at the next order.
The resulting `resonant subgraph' is therefore locally tree-like and we argue that the statistical properties of the resulting eigenstates in the localized phase can be understood via an associated bond percolation problem on the hypercube.

This mapping however does not capture the transition region, in particular, the heterogeneity in single-site observables.
This is not particularly surprising as we have neglected higher-order processes.
Ref.~\cite{Altshuler:1997aa} treats these processes within the tree approximation in a related model.
Their arguments seem to predict an intermediate delocalized non-ergodic phase between $J_c(N) \sim W/(N\log N)$ and $J^*(N) \sim W/N$, that is, between the MBL phase and the thermal phase in the central spin model.
The numerical data at the accessible system sizes is not conclusive about the existence of this possible intermediate phase or the logarithmic suppression of the critical coupling.
We speculate on the possible phase diagrams in the thermodynamic limit in Sec.~\ref{sec:pt_higher_order}.

In what follows, we first describe the model in Sec.~\ref{sec:model} and present the numerical diagonalization results in Sec.~\ref{sec:numerics}.
We then turn to the perturbative analysis at low orders in Sec.~\ref{sub:pt_first_order} and ~\ref{sub:pt_second_order} and compare their quantitative predictions for the MBL phase and the crossover region to the numerical results in Sec.~\ref{sub:pt_consequences}.
Finally, we discuss the role of higher order processes in Sec.~\ref{sec:pt_higher_order}.

\section{Model}
\label{sec:model}

Our Hamiltonian for a thermal inclusion connected to $N$ l-bits is:
\begin{align}
H &= \sum_i \tilde{\Delta}_i \tau_i^z + J \sum_i (A_i  \tau^z_i + B_i  \tau^x_i)
\label{Eq:TheHamiltonian}
\end{align}
where $\tilde{\Delta}_i$ are independently sampled random variables drawn from the box distribution $[-W, W]$, $\tau^\alpha_i$ for $\alpha=x,y,z$ are the Pauli operators of l-bit $i$, and $A_i$ and $B_i$ are real Gaussian symmetric random matrices (GOE) acting on the thermal inclusion Hilbert space of dimension $2^M$.
The normalization of the GOE matrices is such that the off-diagonal elements have variance $1/2^M$, while diagonal elements have twice the variance. This normalization guarantees that the operator norm of $A_i, B_i$ is order 1, which is the appropriate scaling for a local operator acting on the inclusion.

In this paper, we focus on the simplest case of \emph{a central spin model} with $M=1$, as it captures most of the physics of the general $M>1$ model in Eq.~\eqref{Eq:TheHamiltonian}.
At $M=1$, it is convenient to expand the $A_i$ matrices in the Pauli basis $\sigma$ of the central spin:
\begin{align}
A_i &= a^0_i  + a^x_i \sigma^x + a^y_i \sigma^y + a^z_i \sigma^z
\label{Eq:GOEInPauli}
\end{align}
As $A_i$ is a GOE matrix, $a^y_i = 0$, while $a^\alpha_i$ for $\alpha=0,x,z$ are independent Gaussian random variables with zero mean and variance $1/2$.
The Hamiltonian for the central spin model is therefore:
\begin{align}
H &= \sum_{i=1}^N J(\vec{a}_i \cdot \vec{\sigma})\tau^z_i + J B_i \tau^x_i
+\Delta_i \tau^z_i ,
\label{Eq:TheHamiltonianM1}
\end{align}
where $\Delta_i = \tilde{\Delta}_i + J a^0_i$ is a renormalized field.

There are two dimensionless scales in the model: the temperature and the coupling constant $J$ in units of $W$.
As we focus on infinite temperature in this article, $J/W$ determines the entire dynamical phase diagram.
At $J=0$, none of the spins are coupled and the eigenstates are trivially localized product states with $\tau^z_i = \pm 1$.
Each $\tau^z$ state of the l-bits is doubly degenerate as the energy is independent of the state of the central spin $\sigma$.
In anticipation of the leading interaction term at non-zero $J$, we choose to work in a basis in which $\sigma$ points along or against the effective $\tau^z_i$-dependent field:
\begin{align}
\vec{h}(\tau) = J  \sum_i \vec{a}_i\tau_i^z
\end{align}
Thus, $\ket{s \hat{h}, \tau}$ for $s = \pm 1$ is an eigenstate of Eq.~\eqref{Eq:TheHamiltonianM1} with $J=0$, and, in fact, even for $J\ne  0$ for $B_i = 0$.
This set of `Fock states' are naturally viewed as the vertices of a $(N+1)$-dimensional hypercube.
We use the shorthand $\tau = (\tau^z_1,\ldots, \tau^z_i,\ldots, \tau^z_N)$ and drop the explicit dependence of $\vec{h}$ on $\tau$ where it is clear from context.

The statistics of the effective field $\vec{h}$ play an important role in the dynamics of the model.
As the sum of $N$ independent Gaussian random vectors of $O(J)$ in the $xz$ plane, $\vec{h}$ has mean zero and typical (root-mean-square) length $J\sqrt{N}$.
Its distribution is Gaussian with respect to disorder fluctuations at fixed $\tau^z$ (by construction) and with respect to varying $\tau^z_i$ in a fixed sample (by the central limiting theorem).

In the opposite limit of $J \gg W$, the bare random fields $\tilde{\Delta}_i$ become negligible.
Each l-bit feels a random field of order $J$ in the $xz$ plane and interacts with the central spin with strength order $J$.
As the disorder and the interactions are of comparable strength, we expect the system to thermalize as $N \to \infty$.
Both the perturbative (in $B$) arguments and numerical exact diagonalization study discussed below confirm this expectation.
However, as $W$ drops out for large $J$, we cannot simply increase $J$ in order to approach a more robust thermalizing limit at finite size.
This unfortunately results in large finite-size effects in the thermalizing regime.

%%%%%%%%%%%%%%%%%%%%%%%%%%%%%%%%%%%%%%%%%%%%%%%%%%%%%%%%%%%%%%%%%%%%%%
\section{Numerical results}
\label{sec:numerics}

\begin{figure}[tb]
\includegraphics[width=\columnwidth]{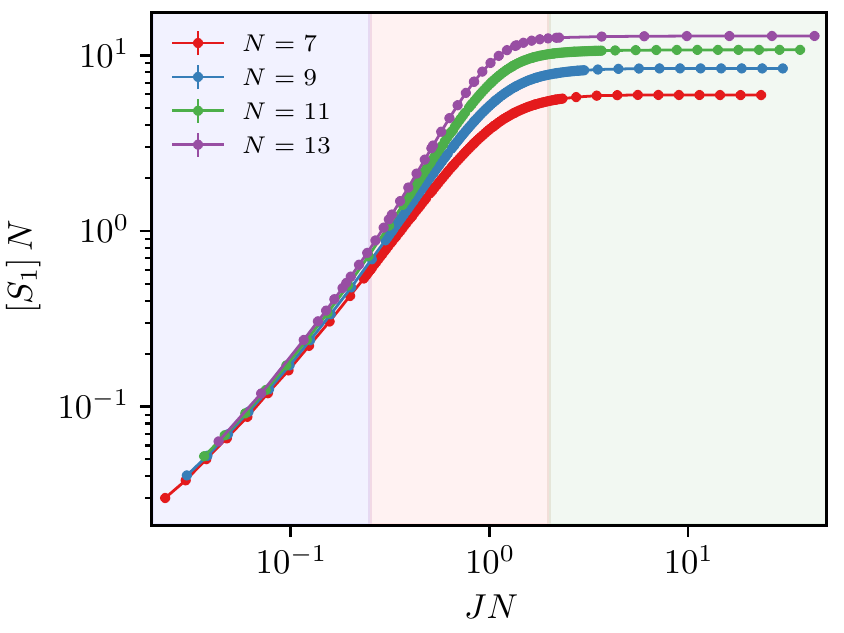}
\caption{Scaled mean single l-bit entanglement entropy $[S_1]N$ vs the scaled interaction strength $JN$ for various $N$.
There are three qualitative regimes:
small $JN \lesssim 0.2$, where the curves collapse, consistent with MBL;
large $JN \gtrsim 1$, where $N$ dependent plateaus form, consistent with the approach to thermalization;
and a wide crossover regime at intermediate $JN$.
}
\label{fig:S1N_JN_loglog}
\end{figure}

We study Eq.~\eqref{Eq:TheHamiltonianM1} using full diagonalization for a $M=1$ central spin coupled to $N = 7$ to $N=13$ l-bits.
The number of samples at each $(N,J)$ is $2500$, except for $N=13$ where this number is $600$, and within each sample, we restrict our analysis to the eigenstates within the central half of the energy spectrum in order to study the properties of infinite temperature.
The mean of a quantity $q$ is denoted by $[q]$, while the standard deviation is denoted by $\Delta q$.
Unless specified otherwise, the mean and standard deviation are taken with respect to all the l-bits (l), the eigenstates in the central half of the spectrum (E), and all samples (s).
When the statistical operation is restricted to a particular subset of these, we include the subset in the subscript.
We measure energy in units where $W=1$.

\subsection{Three Regimes} % (fold)
\label{sub:numerics_three_regimes}

Numerically, we find three distinct regimes as a function of rescaled coupling $JN$: localized (MBL) at small $JN$, thermal (ETH) at sufficiently large $JN$ and a wide crossover at intermediate $JN$.
This is conveniently summarized by the behavior of the mean single l-bit entanglement entropy $[S_1]$ (see Fig.~\ref{fig:S1N_JN_loglog}).
The entropy, $S_1^i$, measures the degree to which l-bit $i$ is thermalized within an eigenstate $\ket{E}$:
\begin{align}
	S_1^i = - \Tr \rho_i \log_2 \rho_i
\end{align}
Here, $\rho_i = \Tr_{\bar{i}} \ket{E}\bra{E}$ is the reduced density matrix for site $i$.
If the state $\ket{E}$ is thermal (at infinite temperature) for l-bit $i$, then $S_1^i = 1$ obtains its maximal value.
In a localized state, on the other hand, $S_1^i$ can be $< 1$.

The  entropy shows the three qualitatively distinct regimes in Fig.~\ref{fig:S1N_JN_loglog}.
At sufficiently small coupling, we find that $[S_1]N$ collapses onto a single curve to excellent precision.
This strongly localized behavior ($[S_1] \sim 1/N \to 0$ as $N \to \infty$) also follows in the perturbative treatment of the localized phase in Sec.~\ref{sec:perturbative_analysis}.
We consider this regime the finite-size precursor to the MBL phase.
At sufficiently large coupling $JN \gtrsim 1$, $[S_1]N$ develops plateaus which increase with $N$, consistent with thermalizing behavior in the thermodynamic limit ($[S_1] \to 1$ as $N \to \infty$).
Finally, the wide crossover between these two behaviors exhibits growth of $[S_1]N$ with $N$, which suggests at least partial delocalization.

\begin{figure}[tb]
\includegraphics{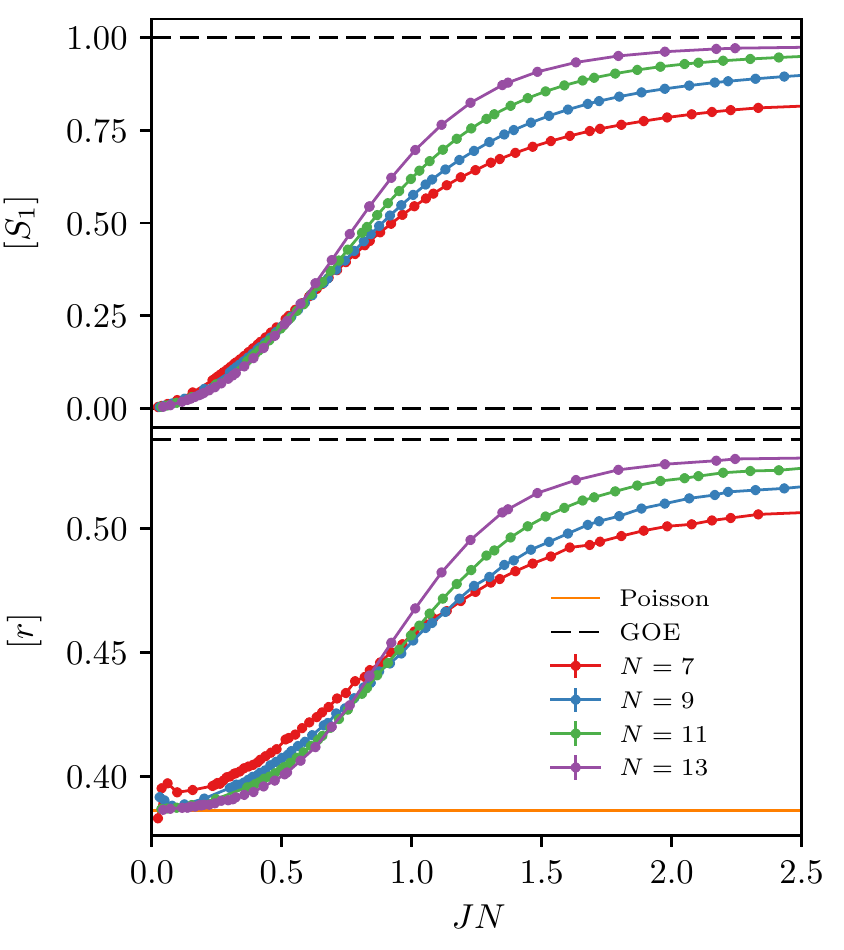}
\caption{(top) Mean single l-bit entanglement entropy $[S_1]$ vs the scaled interaction strength $JN$ and,
(bottom) mean level spacing ratio $[r]$ vs $JN$, at different $N$.
Both plots show sharpening of the transition as $N$ is increased, but with
significant finite-size ``drift'' of the crossing points, particularly for $[r]$.
}
\label{Fig:S1_R_JN}
\end{figure}

In order to focus in on the intermediate transition regime, we look at the finite-size behavior of
$[S_1]$ (unscaled) along with that of the mean level spacing ratio $[r]$ (see Fig.~\ref{Fig:S1_R_JN})
\footnote{The level spacing ratio $r$ is defined as the ratio of consecutive level spacings $r(n) = \textrm{min}(\delta(n),\delta(n+1))/\textrm{max}(\delta(n),\delta(n+1))$ with $\delta(n) = E_n - E_{n-1}$ when the energies $E_i$ are enumerated in increasing order.}.
The ratio measures the level repulsion in a system and is commonly used to diagnose (de)-localization.
It flows to $r_{\textrm{GOE}} \approx 0.53$ in systems with random matrix level statistics and to $r_{\textrm{Poisson}} \approx 0.39$ for systems with Poisson level statistics \cite{Oganesyan:2007aa,Atas:2013aa}.
The upper panel of Fig.~\ref{Fig:S1_R_JN} indicates that $[S_1]$ has a sharpening crossover from $0$ to $1$ near $JN \sim 0.5$ while the lower panel shows a similar sharpening crossover in $[r]$ at $JN \sim 1$.
This suggests that the crossover between the localized and thermal phases sharpens into a phase transition on the scale $JN$.
However, as the finite-size effects are large, it is difficult to separate two possible scenarios:
The first scenario posits that the two crossover points coalesce as $N \to \infty$ and there is a direct transition from MBL to a fully delocalized thermal phase on the scale $JN$ (up to multiplicative logarithmic corrections).
In the second scenario, the crossover points remain separate as $N \to \infty$, sharpening into two phase transitions surrounded an intervening partially delocalized phase \cite{Altshuler:1997aa}, where $[S_1]=1$, the level statistics are Poisson, and where the off-diagonal matrix elements of local operators would presumably not satisfy ETH.
This is often referred to as a delocalized non-ergodic phase.

We comment that $[S_1]$ and $[r]$ both attain plateaus at large $JN$ in the thermal phase. We have checked that the plateau values approach their limiting ETH values as $2^{-N/2}$.

\subsection{Heterogeneity in Observables} % (fold)
\label{sub:numerics_observables}

\begin{figure}[tb]
	\centering
	\includegraphics[width=\columnwidth]{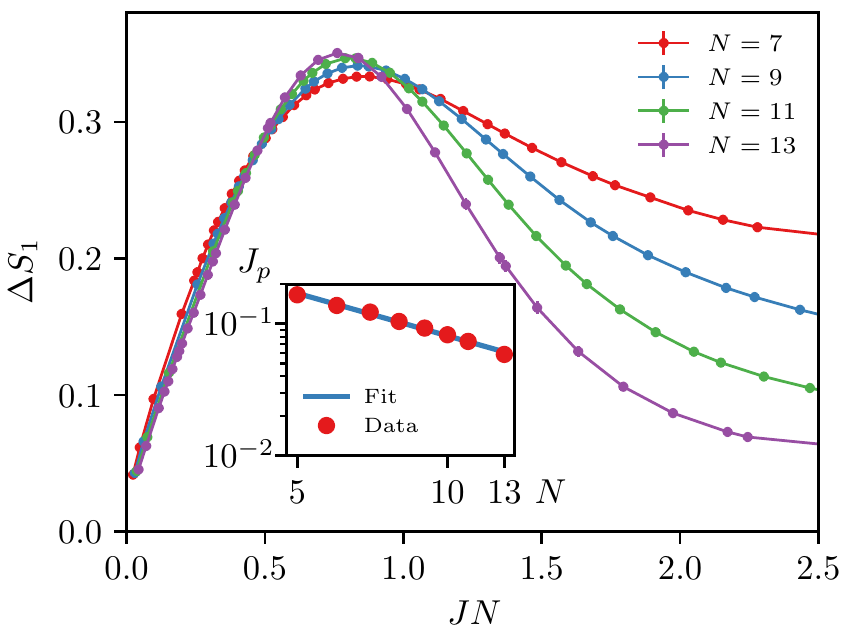}
	\caption{
	Standard deviation of $S_1$ over l-bits, states and samples $\Delta S_1$ vs scaled coupling $JN$.
	(inset) Scaling of peak position $J_p$ with $N$ on log-log scale. Least squares fit (blue) to $J_p \approx 0.95 N^{-1.07}$.
	}
	\label{fig:DeltaS_JN}
\end{figure}

The most striking feature of the crossover regime is the extreme heterogeneity of single site observables across eigenstates, samples and l-bits.
As a coarse measure, consider the standard deviation of the single site entanglement entropy $\Delta S_1$ shown in Fig.~\ref{fig:DeltaS_JN}.
In both the MBL and ETH regimes of Fig.~\ref{fig:DeltaS_JN}, the ``flow'' of $\Delta S_1$ is to 0 as $N$ grows.
This is to be expected in the infinite temperature ETH phase, where the fluctuations in single-site observables across eigenstates are exponentially small in $N$ ($\Delta S_1 \sim 2^{-N/2}$ to be precise).
On the MBL side, the perturbative picture of Sec.~\ref{sec:perturbative_analysis} indicates that the entire distribution of $S_1$ scales to $0$ as $1/N$ so that $\Delta S_1 \sim 1/N$ as well~\footnote{In contrast, in the one dimensional MBL phase, $\Delta S_1$ is non-zero as $N \to \infty$.}.
The data in Fig.~\ref{fig:DeltaS_JN} are consistent with these predictions.

In the crossover regime, however, $\Delta S_1$ shows a peak whose height $\Delta S_1^p$ increases with $N$.
As $S_1$ is a bounded variable, the growth of $\Delta S_1^p$ must saturate at larger sizes.
Nonetheless, should the peak persist to the thermodynamic limit, it must converge to a critical point (see the discussion below).
Previous works in one dimensional models have likewise used the peak as a sensitive proxy for the critical point \cite{Kjall:2014aa,Khemani:2016aa}.
We accordingly define $J_p(N)$ by the location of the peak in $\Delta S_1$ and study the properties of the critical region by following this coupling.
The inset to Fig.~\ref{fig:DeltaS_JN} confirms that it scales as $J_p(N) \sim W/N$.

\begin{figure}[tb]
\includegraphics[width=\columnwidth]{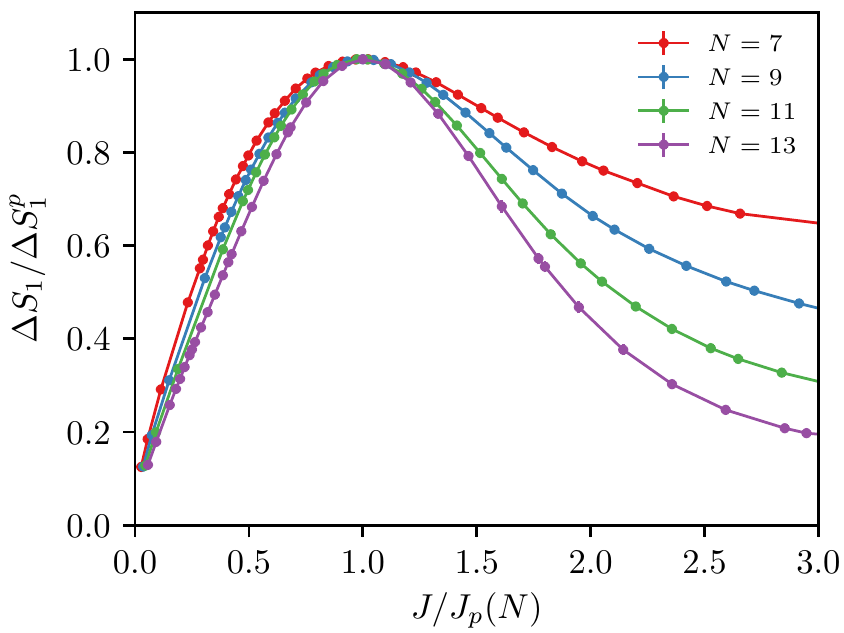}
\caption{ $\Delta S_1 /\Delta S_1^{p}$ narrows on the scale $J/J_p(N)$ with increasing system size. This is consistent with the finite-size crossover sharpening into a delocalization phase transition in the limit of large $N$.}
\label{fig:Rescaled_DeltaS_Peak}
\end{figure}

\begin{figure}
\includegraphics[width=0.9\columnwidth]{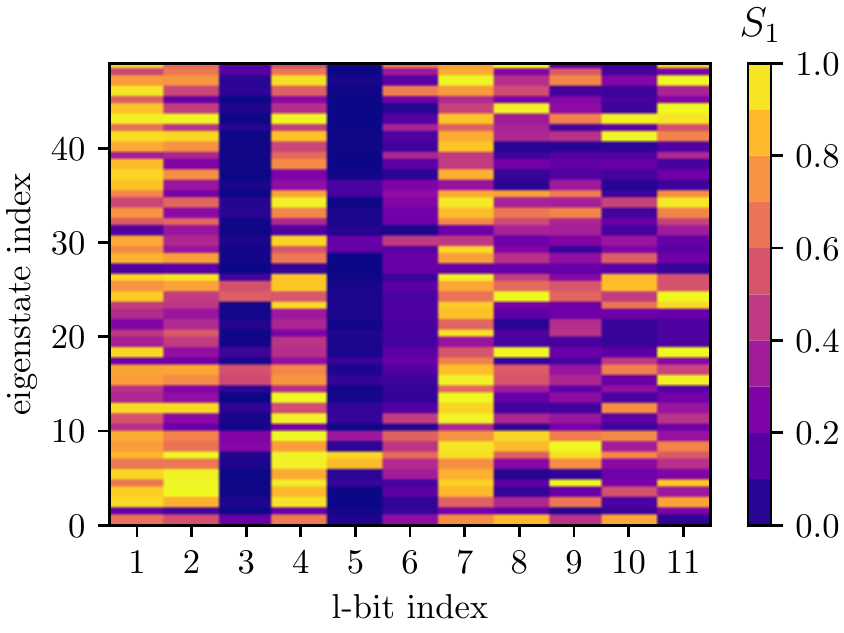}
\caption{ $S_1$ heatmap from a specific sample of the central spin model with $N=11$ l-bits at the peak coupling $J_p$. Each row corresponds to an eigenstate from 50 infinite temperature selected eigenstates and each column to a particular l-bit.}
\label{fig:Sample_S1_snapshot}
\end{figure}

\begin{figure*}[ht]
\includegraphics[width=0.329\textwidth]{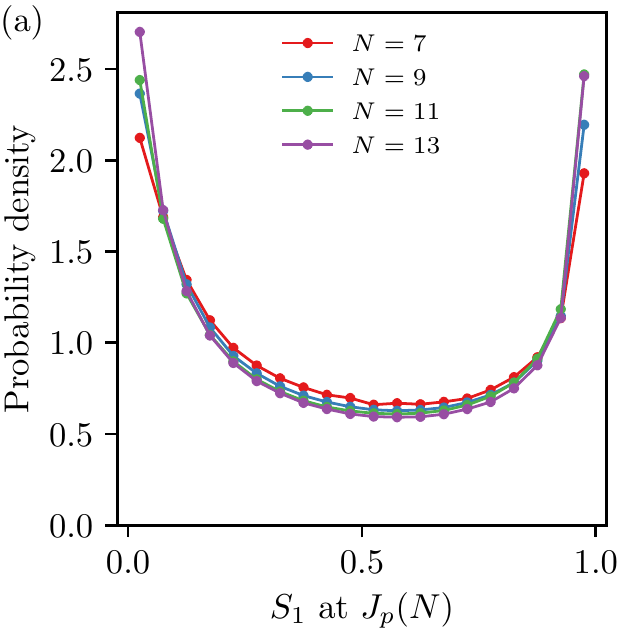}
\includegraphics[width=0.329\textwidth]{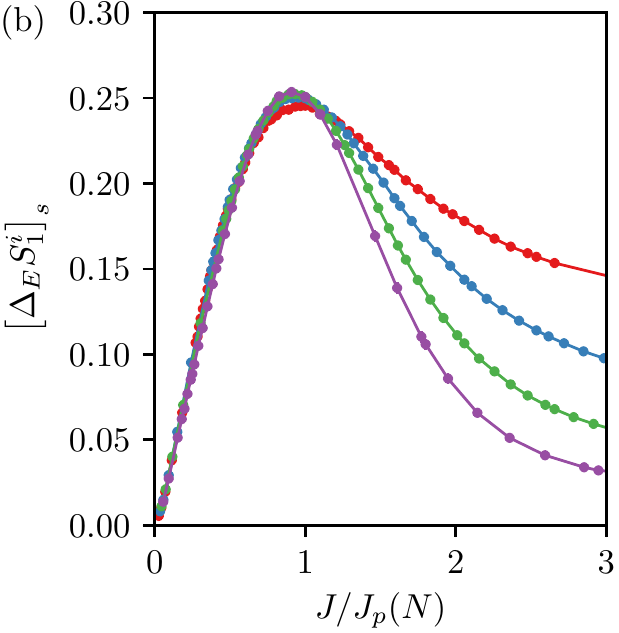}
\includegraphics[width=0.329\textwidth]{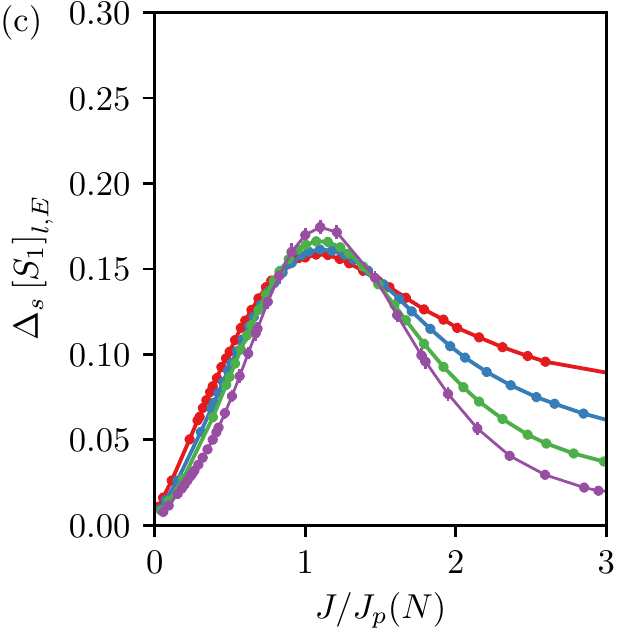}
\caption{
 (a) Histogram of $S_1$ collected across l-bits, states and samples at $J_p$. (b) Standard deviation of $S_1^i$ for a fixed l-bit $i$ among eigenstates from the same sample and averaged over different samples $\left[\Delta_E S_1^{i}\right]_{s}$ vs scaled $J$.  (c) Standard deviation over samples of $S_1$ averaged within each sample over eigenstates and l-bits vs scaled $J$.}
\label{fig:S1_bimodality}
\end{figure*}

The relative width of the peak at $J=J_p(N)$ narrows as $N$ increases (see Fig.~\ref{fig:Rescaled_DeltaS_Peak}).
This trend is consistent with the existence of a sharp phase transition in $S_1$ at $J_p(N)$ in the thermodynamic limit.
Indeed, $J=J_p(N)$ approaches the location of the crossing point in $[S_1]$ in Fig.~\ref{Fig:S1_R_JN}(a) at $JN\approx 0.5$ with increasing $N$.
As $S_1$ detects if l-bits are entangled in eigenstates, we expect that it is sensitive to whether the system is localized or delocalized on the classical hypercube of `Fock states', but not necessarily to whether the delocalized phase satisfies the off-diagonal criteria for ETH.
Fig.~\ref{fig:Rescaled_DeltaS_Peak} is therefore consistent with both scenarios for the phase diagram discussed in Sec.~\ref{sub:numerics_three_regimes}.
In the first scenario, there is a direct phase transition from a MBL to an ETH phase at $J=J_p(N)$ as $N \to \infty$.
In the second scenario in where there are two phase transitions with an intervening partially delocalized phase, $J = J_p(N)$ is the position of the first transition out of the localized phase.

We note that the narrowing of this peak on the localized side is slower than on the delocalized side.
This makes the extraction of a finite-size scaling exponent problematic, and indicates that the numerics are not yet in the asymptotic scaling regime as we discuss below.
Moreover, the narrowing on the delocalized side suggests exponents which violate finite size scaling bounds \cite{Chayes:1986kq,Chandran:2015aa}.

We now characterize the heterogeneity in $S_1$ in the crossover region in more detail; see also Appendix~\ref{app:std_dev_parsing}.
Fig.~\ref{fig:Sample_S1_snapshot} shows the wide variation of $S_1$ across both l-bits and eigenstates within a typical sample.
More quantitatively, Fig.~\ref{fig:S1_bimodality}(a) shows that the distribution of $S_1$ across l-bits, eigenstates and samples at $J=J_p(N)$ is increasingly bi-modal with increasing $N$, with increasing weight near zero and one bit and decreasing weight at intermediate values of $S_1$.
Thus, l-bits are mostly either `on' (highly entangled) or `off' in any given eigenstate in the crossover regime.
The bi-modal distribution leads to the large peak in $\Delta S_1$ seen in Fig.~\ref{fig:DeltaS_JN}.

As l-bits in weak (strong) fields might strongly (weakly) entangle with the central spin and with one another, the reader may not be surprised by the bi-modality seen at $J_p$.
This explanation however misses the remarkable feature that \emph{different} l-bits are active in different eigenstates.
To demonstrate this effect, we investigate the sample averaged variation of $S_1^i$ across eigenstates, $\left[\Delta_E S_1^{i}\right]_{s}$, for fixed l-bit $i$ (see Fig.~\ref{fig:S1_bimodality}(b)).
If l-bit $i$ were entangled/unentangled across all the states, $\left[\Delta_E S_1^{i}\right]_{s}$ would be small.
Instead, we see a robust peak in this quantity at $J=J_p(N)$ at each $N$, with a magnitude comparable to the total variation of $S_1$ across all samples, states and l-bits ($\Delta S_1$).

In Appendix~\ref{app:std_dev_parsing}, we analyze the contributions to $\Delta S_1$ coming from sample, eigenstate and l-bit fluctuations, all of which contribute significantly to the total variation at these sizes.
In Fig.~\ref{fig:S1_bimodality}(c) we single out the sample-to-sample variation, as this is the only component which shows an {\it accelerating} trend to stronger and sharper peaks as $N$ increases.
This is another indication that we are not in the asymptotic scaling regime.
We also note that the inhomogeneity between eigenstates in the l-bit-averaged entanglement at the transition in Fig. \ref{fig:appendix_A_1}(b) decreases with $N$.
If this persists to larger $N$, then all the eigenstates in one sample will be similar in this respect (although it seems likely they will still differ in which l-bits are less versus more entangled).

Finally, we note that this central spin model exhibits the strongest single-site bimodality in the MBL-ETH crossover of any numerically studied model to date, suggesting that the transition may indeed be in some sense `first order' \cite{Yu:2016aa}.
Furthermore, Fig.~\ref{fig:S1_bimodality}(b) shows that $S_1^i$ fluctuates strongly between eigenstates of a single sample, even at infinite temperature.
No RG treatment to date takes these intra-sample variations between eigenstates into account, so might be missing some important physics of this transition \cite{Vosk:2015aa,Potter:2015ab}.

\subsection{Eigenstate Distributions} % (fold)
\label{sub:eigenstate_distributions}

\begin{figure*}[!t]
\includegraphics[width=1.0\textwidth]{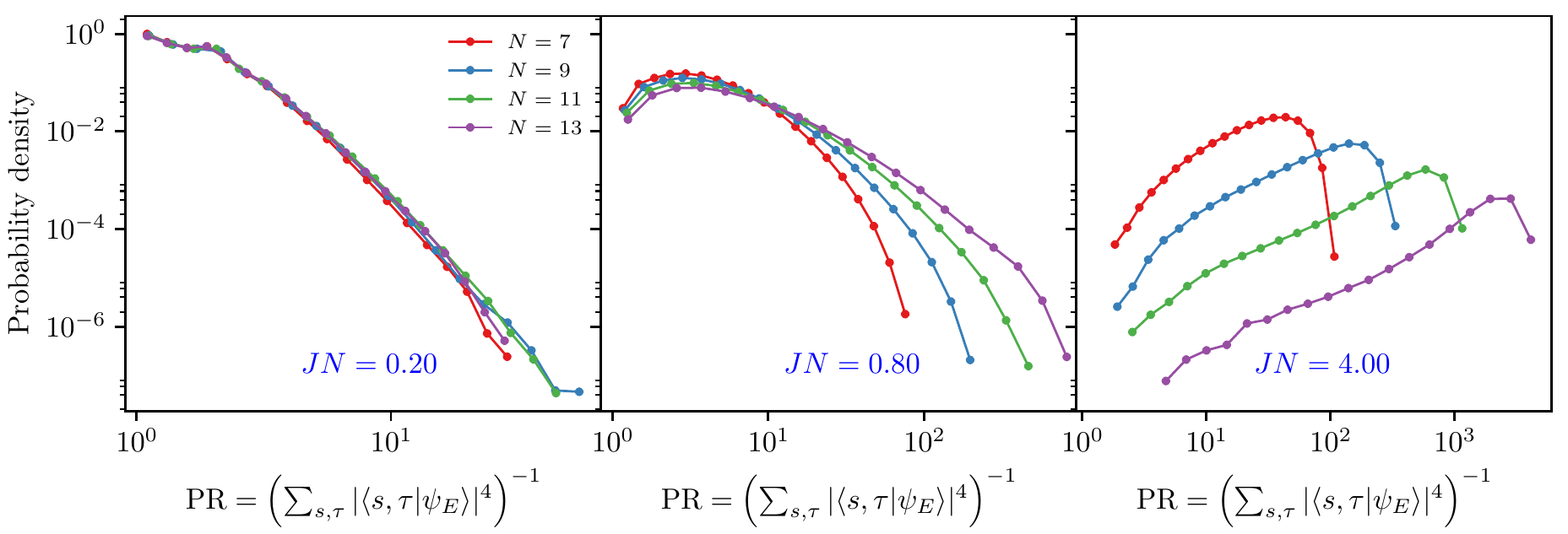}
\caption{Probability density function of the eigenstate participation ratio (PR)
on the classical basis $|s\hat h,\tau\rangle$ for $JN$ in the three different regimes:
deep in the MBL phase ($JN=0.2$), in the crossover region ($JN=0.8$), and
in the thermal phase ($JN=4.0$).
 }
\label{fig:PR_distributions_vs_JN}
\end{figure*}

As discussed in Sec.~\ref{sec:model}, the eigenstates at $JN=0$ are the product states $|s\hat{h}, \tau\rangle$.
It is therefore natural to study the support of the eigenstates at $JN>0$ on this basis (which forms a hypercube).
This is measured by the participation ratio $\textrm{PR}$ of an eigenstate $|E_m\rangle$:
\begin{align}
	\textrm{PR}^{-1}(E_m) = \sum_{s,\tau} |\langle E_m | {s\hat{h},\tau}\rangle |^4
	\label{Eq:ClassicalPR}
\end{align}

Fig.~\ref{fig:PR_distributions_vs_JN} shows the evolution of the distribution of the participation ratios with coupling.
In the localized phase (illustrated by $JN=0.2$), the distribution strongly decays with PR and is independent of $N$.
This implies that the eigenstates are essentially localized to finite regions of the hypercube even in the thermodynamic limit.
This is consistent with the perturbative picture of the localized phase in Sec.~\ref{sec:perturbative_analysis}.

In the crossover regime (at $JN=0.8$), the distribution remains peaked at small PR, but the weight in the tail grows with $N$.
Further, the scale of the cut-off grows exponentially with $N$.
In the second-order perturbative treatment developed in Sec.~\ref{sec:perturbative_analysis}, the transition coincides with the high-dimensional percolative transition on the hypercube.
In this picture, the PR distribution should decay as a power law with exponent $3/2$ which is cutoff by $n_\xi \sim 2^{2N/3}$.
This scaling of the cutoff is in agreement with the numerical data at $JN=0.8$, although the curvature in the bulk of the distribution is clearly inconsistent with a simple power law.

Finally, deep in the thermalizing phase (at $JN=4$), the PR at the peak of the distribution increases exponentially with $N$, consistent with ETH.
The probability density at finite PR appears to decrease exponentially with $N$, consistent with exponentially rare samples probing the localized side of the transition.

%%%%%%%%%%%%%%%%%%%%%%%%%%%%%%%%%%%%%%%%%%%%%%%%%%%%%%%%%%%%%%%%%%%%%%%

\section{Perturbative analysis}
\label{sec:perturbative_analysis}
The central spin model is amenable to perturbative study in small $B$ relative to the ``classical'' configurations $\ket{s \hat{h}, \tau}$ defined in Sec.~\ref{sec:model}.
In this approach, delocalization takes place when a typical starting configuration resonates with a divergent subgraph of degenerate configurations on the hypercube.

In Secs.~\ref{sub:pt_first_order} and \ref{sub:pt_second_order}, we consider first and second order processes and show that a delocalization transition takes place at $J_c \sim W/N$.
At this coupling, only a small (finite) number of spins can resonate by first-order spin-flip processes.
On the other hand, second order `flip-flop' processes, in which two l-bits and the central spin are flipped, produce a divergent subgraph of resonant configurations for $J > J_c$.
Thus, the low order perturbative analysis captures a delocalization transition on a scale $J_c \sim W/N$, consistent with the numerical observations.

As the resonant subgraphs produced at second order are locally tree-like with an expected branching number $K \sim J^2N^2/W^2$, their statistical properties can be determined from independent second neighbor bond percolation on the hypercube with probability of placing a bond, $p\sim J^2/W^2$.
We summarize the relevant properties of this percolation problem in Sec.~\ref{sub:pt_props_resonant_subgraphs}.
In Sec.~\ref{sub:pt_consequences}, assuming that actual many-body eigenstates are simply delocalized over these resonant subgraphs, we explain the numerical observations that $[S_1] \sim 1/N$ in the localized phase, as well as the distribution of the participation ratios in the classical basis.
Thus, the low-order perturbative analysis also quantitatively describes the localized phase for $J < J_c$.

However, the low-order perturbative analysis does not capture all of the features of the crossover region at $ J \approx J_c$.
This is not surprising as we expect higher order processes to be important in the vicinity of the delocalization transition.
In Sec.~\ref{sec:pt_higher_order}, we adapt the arguments of Ref.~\cite{Altshuler:1997aa} that suggest a transition from localized to a delocalized non-ergodic phase on the scale $J \sim W / N \log N$.
As discussed in Sec.~\ref{sec:numerics}, the numerical evidence for this intervening phase is inconclusive.

\subsection{Preliminaries}
\label{sub:pt_preliminaries}

At $B=0$, the state $|s\hat{h}, \tau\rangle$ has classical energy
\begin{align}
	E_{s\tau} = s|\vec{h}(\tau)| + \sum_{i=1}^N \Delta_i \tau^z_i \label{Eq:ClassicalEnergy}
\end{align}
These $2^{(N+1)}$ configurations are naturally viewed as the corners of a hypercube.
The interaction term, $J \sum_i B_i \tau^x_i$, defines a short range hopping model on the hypercube as it connects the corners related by single l-bit flips either with or without flipping the central spin $s$.
As the perturbative arguments will predict a delocalization transition at $J_c \sim W/N \ll W$, we assume $J \ll W$ henceforth, since here we are only considering the large-$N$ regime.

To perturb around a typical infinite-temperature configuration $\ket{s \hat{h}, \tau}$, we need the classical energies of nearby configurations.
We denote the l-bit configuration obtained by flipping $k$ l-bits $i_1, \ldots, i_k$ by $\tau^\prime_{i_1 \ldots i_k} = (\tau^z_1, \ldots, -\tau^z_{i_1},\ldots, -\tau^z_{i_2},  \ldots, \tau^z_N)$.
From Eq.~\eqref{Eq:ClassicalEnergy}, the energy of such a state relative to the initial state is
\begin{align}
\label{eq:energydiffexact}
	\Delta E_{s' \tau'_{i_1 \ldots i_k}} &=  s' |\vec{h}(\tau')| - s |\vec{h}(\tau)| - 2 \sum_{m=1}^k \tau^z_{i_m} \Delta_{i_m}
\end{align}
For small numbers of flipped l-bits ($k \ll N$), the deviation in the effective field on the central spin,
\begin{align}
	J \vec{\delta}_{i_1, \ldots i_M} &\equiv \vec{h}(\tau')-\vec{h}(\tau) = -2 \sum_{m=1}^k \vec{a}_{i_m} \tau^z_{i_m}
\end{align}
is much smaller than the field itself, $|\vec{\delta}_{i_1, \ldots i_M}| \sim \sqrt{k} \ll \sqrt{N}$.
Thus, expanding Eq.~\eqref{eq:energydiffexact} in $J \vec{\delta}/|\vec{h}(\tau)|$ (and suppressing the indices of the $k$ flipped l-bits),
\begin{align}
\label{eq:classenergydifferences}
	\Delta E_{s' \tau'} &=(s' - s)|\vec{h}|% \nonumber\\
	- 2 \sum_{m=1}^k \tau^z_{i_m}(\Delta_{i_m} 	+ s' J \vec{a}_{i_m} \cdot \hat{h}) \nonumber\\
	&+ s' \frac{J^2}{2|\vec{h}|}\left( |\vec{\delta}|^2 - (\hat{h}\cdot \vec{\delta})^2 \right) + O\left(\frac{J^3}{|\vec{h}|^2}\right)
\end{align}
The first term is the dominant energy change on flipping the central spin relative to its local field. It is of order $J\sqrt{N}$ but independent of the choice of flipped l-bits.
The second term is the total field energy of the flipped l-bits. The bare fields $\Delta_i \sim W$ while the correction due to the central spin is only $\sim J \ll W$.
Finally, the third and higher order terms represent the interactions between the flipped l-bits induced by their interaction with the central spin.
In particular, the leading interaction is of order $J/\sqrt{N}$.

\subsection{First order processes} % (fold)
\label{sub:pt_first_order}

At first order in $B$, there are $2N$ neighboring configurations with a single l-bit flipped with or without flipping the central spin. The matrix element,
\begin{align}
	J\langle s' \hat{h}(\tau'_j)|B_j|s\hat{h}(\tau)\rangle \sim O(J)
\end{align}
while the energy differences $\Delta E_{s'\tau'_{j}}$ are distributed on a band of width $W$.
As the level spacing is $\sim W/2N$, there are $\sim 2 N J / W$ resonant neighbors.
Thus, first order processes begin to find $O(1)$ resonances when $J \sim W/N$.
We call these `step 1' first order resonances.

Diagonalizing the perturbation exactly on these resonant configurations produces new basis states delocalized over the resonant cluster.
These in turn connect at first order to configurations on the hypercube with two l-bit flips.
For $J \sim W/N$, there are again $O(1)$ next neighbor states which satisfy the resonance condition (`step 2' resonances).
However, as the interaction energy in Eq.~\eqref{eq:classenergydifferences} $\sim J/\sqrt{N} \sim W / N^{3/2}$ is much smaller than the level spacing $W/N$, roughly the same l-bit flips that were resonant at step 1 are resonant at step 2.

\begin{figure}[htb]
	\centering
	\includegraphics[width=\columnwidth]{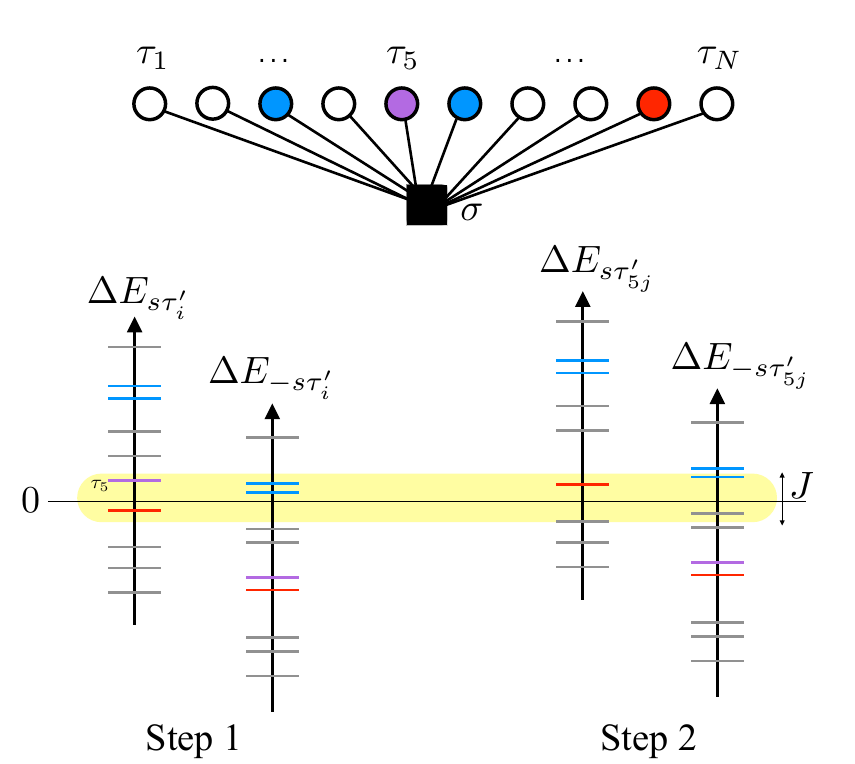}
	\caption{
	(top) Illustration of $N$ l-bits coupled to the central spin $\vec{\sigma}$.
	(bottom)
	Schematic of first order processes contributing to resonance relative to reference configuration $s \tau$.
	At step 1, there are $N$ single l-bit flipped states with central spin $s$ and $N$ with central spin $-s$. Flipping the red/purple l-bits leads to red/purple states which satisfy the resonance condition $|\Delta E_{s \tau'}| < J$ without flipping the central spin; similarly, for the blue l-bits with the central spin flipped ($-s$).
	Step 2 illustrates the levels accessible from the purple state ($\tau_5$ flipped) at step 1. As the energy shifts are very small, roughly the same l-bit flips are resonant.
	}
	\label{fig:firstorderschematic}
\end{figure}

More precisely, if l-bit flip $i_1$ is resonant at step 1, $\Delta E_{s'\tau'_{i_1}} < O(J)$, then l-bit $i_2$ can resonantly flip at step 2 only if $\Delta E_{s''\tau'_{i_1i_2}} < O(J)$.
This requires that the effective field on l-bit $i_2$, $2\left(\Delta_{i_2} + s'' J \vec{a_{i_2}}\cdot \hat{h}\right)$ itself should be close to $0$ (if $s''=s$) or $|\vec{h}|$ (if $s''=-s$) to an accuracy of $O(J)$.
The interaction energy between l-bits $i_1$ and $i_2$ does not change the above resonance condition for l-bit $i_2$ at step 2 because it is negligible on the energy scales $J \approx W/N$.
As there are only $O(1)$ l-bits whose field energy are close to $0$ or $|\vec{h}|$ to accuracy $O(J)$, even multi-step first order processes only produce a small cluster of resonant configurations. See Fig.~\ref{fig:firstorderschematic}.

We have neglected the $O(J)$ shifts in the reference energy $E_{s\tau}$ which arise due to the diagonalization over the resonant configurations. This is analogous to neglecting the real part of the self energy in a locator expansion.
As these corrections effectively shift the resonance condition rigidly for all l-bits, they do not modify the statistics of the resonant clusters generated.

The structure of the resonant clusters suggests that there will be eigenstate dependence of which l-bits resonantly flip at $J \sim W/N$.
There is an $O(1)$ subset of the l-bits with fields $\Delta_i \sim O(J)$ which resonate across all eigenstates without flipping the central spin.
Meanwhile, those $\sqrt{N}$ l-bits with larger fields $\Delta_i \sim J \sqrt{N}$  resonate (in $O(1)$ groups) only in those eigenstates whose central spin field matches $\Delta_i$.

\subsection{Second order processes} % (fold)
\label{sub:pt_second_order}

As all but $O(1)$ of the $2N$ nearest neighbor states are off-resonant when $J \sim W/N$, in the following we neglect the first order resonances entirely.
At second-order in $B$, there are $2\binom{N}{2} \sim N^2$ neighboring configurations with two l-bits flipped with or without flipping the central spin.
In Appendix~\ref{app:pt_second_order_details}, we show that the effective matrix elements are largest to the subset of states $|-s\hat{h}(\tau'_{ij}), \tau'_{ij}\rangle$ in which the central spin is flipped from $s$ to $-s$ relative to the central field.
This is physically sensible as delocalization in this model proceeds due to the interactions with the central spin, so that the divergent resonant subgraph should include classical configurations in which the central spin is flipped.
The effective matrix elements to the states $|-s\hat{h}(\tau'_{ij}), \tau'_{ij}\rangle$ are typically $\sim J^2/W$.
As there are $\sim N^2$ such final states with energies $\Delta E_{-s \tau'_{ij}}$ spread over a bandwidth $W$, the level spacing is $\sim W/N^2$.
This leads to $\frac{J^2}{W} / \frac{W}{N^2}$ resonances, so that second order processes begin to find $O(1)$ resonances when $J \sim W/N$ (`step 1 second order' resonances).

Now consider the $\sim  N^2$ states accessible from a step 1 resonant state with two more l-bits flipped and the central spin returned to $s$ (as the central spin flips at each step).
In order for the l-bit pair $kl$ to be resonant, the state must have energy $|\Delta E_{s\tau'_{ijkl}}| < \frac{J^2}{W}$.
This is on the scale of the level spacing $W/N^2$.
As $\Delta E_{s\tau'_{ijkl}}$ does not contain the energy change on flipping the central spin $\sim J \sqrt{N}$ and $J\sqrt{N}$ is much greater than the level spacing $W/N^2$, the pairs $kl$ that satisfy the resonance condition at step 2 are typically not the same as the pairs $ij$ that satisfy the resonance condition at step 1.
Thus, the resonant subgraph includes new l-bit flips at step 2 as compared to step 1.

At step 3, the $\sim N^2$ accessible states involve three l-bit pair flip-flops with the central spin again flipped to $-s$.
The resonance condition $|\Delta E_{-s \tau'_{ijklmn}}| < \frac{J^2}{W}$ picks \emph{different} pairs as compared to step 1 because the interaction energy,
\begin{align}
\label{eq:intenergy}
	E^{int} &= -s \frac{J^2}{2|\vec{h}|}\left( |\vec{\delta}|^2 - (\hat{h}\cdot \vec{\delta})^2 \right) \\
	\vec{\delta} &= -2 (\tau^z_i \vec{a}_i + \tau^z_j \vec{a}_j  + \tau^z_k \vec{a}_k + \tau^z_l \vec{a}_l + \tau^z_m \vec{a}_m + \tau^z_n \vec{a}_n) \nonumber
\end{align}
has cross terms between the pairs $ij$, $kl$ and $mn$ which cannot be ignored.
The cross terms are typically $\sim J^2/|\vec{h}| \sim W /N^{3/2}$ which is much larger than the level spacing $W/N^2$.
Thus,  the pairs $mn$ that satisfy the resonance condition $|\Delta E_{-s\tau'_{ijklmn}}| < J^2/W$ are not the same as the pairs $ij$ that satisfy the resonance condition $|\Delta E_{-s\tau'_{ij}}| < J^2/W$ at step 1.
When $J\sim W/N$, we therefore find $O(1)$ \emph{completely new} resonant states among the configurations accessible by flip-flops at \emph{each} step.

\begin{figure}[htb]
	\centering
	\includegraphics[width=\columnwidth]{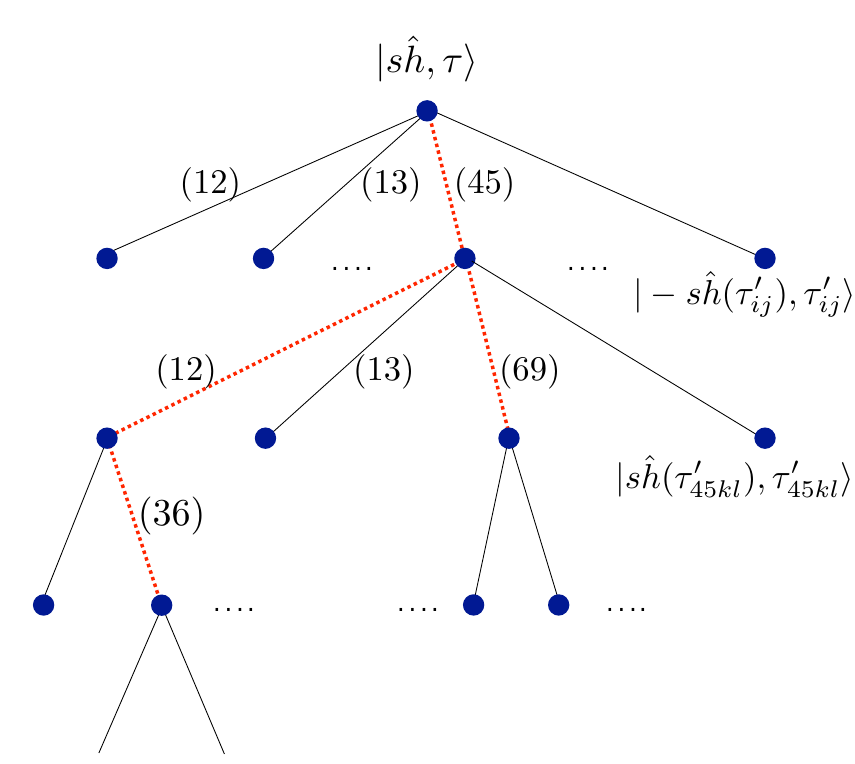}
	\caption{Schematic of graph generated by second order processes on the hypercube of classical configurations.
	At each step, there are $Z \sim N^2$ pairs that may be flipped (edges) of which only $K \sim O(1)$ are resonant (red dashed).
	The shown resonant subgraph has $n=5$ states.
	}
	\label{fig:secondordertree}
\end{figure}

From the above construction, we see that the resonant subgraph of the hypercube generated by second order processes is tree-like for small enough $J$, see Fig.~\ref{fig:secondordertree}.
The expected branching number of the tree is $K \sim J^2N^2/W^2$.
As is well known, such random trees undergo a continuous percolation transition at $K = K_c \sim 1$ at which the probability that the resonant subgraph is infinite starts growing from zero \footnote{When the expected branching number $K$ becomes too large, we expect the tree approximation to break down (as short loops in the interaction graph become very important).  As the critical branching number $K_c$ is of order one, our treatment is self-consistent within the MBL phase.}.
This provides a $J_c(N) \sim \sqrt{K_c} W/N$ beyond which second order resonances guarantee that typical eigenstates cannot remain localized on the hypercube.

\subsection{Properties of resonant subgraphs}
\label{sub:pt_props_resonant_subgraphs}

In this section, we model the resonant subgraphs generated by second-order processes by independent bond percolation on the hypercube with second neighbor bonds placed with probability $p \sim J^2/W^2$. The following properties of resonant subgraphs then follow \cite{Borgs:2001fw,Heydenreich:2015aa}:

\begin{enumerate}
\item For $J< J_c$, the number of resonant subgraphs of size $n$ per site in Hilbert space is given by:
\begin{align}
	c(n) \propto n^{-5/2} e^{-n/n_\xi} \label{Eq:ClusterSizeDist}
\end{align}
with
\begin{align}
	n_\xi \sim \left(\frac{J - J_c}{J_c}\right)^{-2}
\end{align}
$n_\xi$ is the characteristic number of sites in a resonant subgraph in the localized phase, which diverges at the transition. At the transition and for $n\ll n_\xi$ in the vicinity of the transition, the subgraphs are `critical' and are distributed according to a power-law.
A randomly chosen site in the hypercube lies in a connected cluster of size $n$ with probability
\begin{align}
\label{eq:sampledClusterDistribution}
	p(n) \propto n c(n) \sim n^{-3/2} e^{-n / n_\xi}
\end{align}

\item
At the critical point $J = J_c$ at finite $N$, we expect Eq.~\eqref{eq:sampledClusterDistribution} to describe the statistics of cluster sizes with a cutoff size $n_\xi$ that diverges with increasing $N$.
Above the upper critical dimension for percolation, the largest cluster scales with the volume of space to the $2/3$ power,
\begin{align}
 	n_\xi \sim V^{2/3} \sim 2^{\frac{2}{3}N}
	\label{Eq:nxiCritical}
\end{align}
This is well-known for the emergence of the giant component in Erd\"os-Renyi random graphs and likewise for percolation on Euclidean lattices of finite size $L$ in dimensions $d > d_u = 6$ \cite{Borgs:2001fw,Bollobas:2001sj}.
Numerical investigation of independent bond percolation on the hypercube (up to $N=20$, not shown) is consistent with this scaling.

Thus, at the critical point at size $N$, the largest resonant subgraphs, while exponentially large in $N$ are still an exponentially small fraction of the hypercube.
We expect that such large subgraphs are equally likely to include states in which l-bit $i$ is up or down.

\item For $J > J_c$, there is a unique giant component which absorbs a fraction $P \propto |J - J_c|^1$ of the total volume $2^N$ of the hypercube. The remaining clusters are of finite size with a distribution $\propto n^{-3/2} e^{- n / n_\xi}$.  In the giant component, all l-bits are equally likely to be up or down, while the remaining clusters have only a finite number of flipped l-bits.

\end{enumerate}

\subsection{Consequences}
\label{sub:pt_consequences}

We now assume that the actual many-body eigenstates are simply delocalized over the resonant subgraphs generated by second-order processes. This leads to several quantitative predictions about the localized phase, which agree with numerical observations in Sec.~\ref{sec:numerics}. On the other hand, the predictions for the crossover region do not explain the numerical data very well and suggest that higher order processes are important.

\begin{enumerate}
\item The transition from localized to delocalized takes place at $J = J_c(N) \sim W/N$.
This is consistent with our numerical observations in Sec.~\ref{sec:numerics}.
However, as we discuss below, higher-order processes may decrease $J_c$ logarithmically in $N$.

\item
In a cluster with $m$ edges, the number of flipped l-bits is at most $2m$ (as at most two distinct l-bits can flip on every edge of the cluster, see Fig.~\ref{fig:secondordertree}).
Thus, the entanglement entropy of only $O(m)$ l-bits can be non-zero in a many-body eigenstate delocalized over this cluster.
In the localized phase, the clusters are finite and thus
\begin{align}
	[S_1] = O(1/N)
\end{align}
This explains the remarkable collapse in Fig.~\ref{fig:S1N_JN_loglog} in the localized phase.
Indeed, the entire distribution of $S_1^i$ has weight only at zero as $N \to \infty$ which implies that $\Delta S_1 \sim 1/N$ in the localized phase.

\item
Within this model, the participation ratio PR of an eigenstate $\ket{E}$ is simply given by the size $n$ of the associated cluster.
Thus, the PR distribution is given by Eq.~\eqref{eq:sampledClusterDistribution}.
This explains several aspects of the numerical PR distributions in Fig.~\ref{fig:PR_distributions_vs_JN}: (1) the PR distribution is nearly $N$-independent for small $JN$ and has a sharp cut-off $n_\xi$, (2) the cut-off scales exponentially with $N$ as predicted by Eq.~\eqref{Eq:nxiCritical} in the crossover region.
However, the power law regimes suggested by this model are not clearly seen in the numerics.

\item
At finite $N$, the localized phase in the vicinity of the transition is predicted to be heterogenous across eigenstates, l-bits and samples as each initial configuration $\tau$ provides a different landscape for building the resonant subgraph.
However, this heterogeneity vanishes as $N \to \infty$ as almost all the l-bits are unentangled in eigenstates (see the discussion of $[S_1]$ above), so this does not explain the numerical observations in the crossover region.

\item
Another possible model of the crossover region is that it is described by the bond percolation problem at $J>J_c$ with a giant component.
For eigenstates delocalized on the giant component, $S_1^i=1$, while for eigenstates delocalized on finite clusters, $S_1^i$ is non-zero only for a finite number of l-bits.
This model therefore predicts that the distribution for $S_1$ becomes bi-modal as $N \to \infty$ with $p(S_1=0) = (1-P)$ and $p(S_1=1)=P$, where $P \sim |J-J_c|$ is the fraction of sites belonging to the giant component.

However, the numerics presented in Sec.~\ref{sec:numerics} do not support this picture.
First, there is no evidence of a giant component in the PR distributions in the crossover region in Fig.~\ref{fig:PR_distributions_vs_JN}(b).
Furthermore, this percolation model does not predict a significant variation of $S_1^i$ between l-bits $i$ in the same state.
Numerically, this is the biggest source of the variation in $S_1^i$ in the crossover region (see Fig.~\ref{fig:appendix_A_1}).

\end{enumerate}

\subsection{Higher order processes}
\label{sec:pt_higher_order}

It is known that the Anderson localization transition for a single particle hopping on a Bethe lattice with large coordination $Z$ takes places at hopping strength $J \sim W/Z\log Z$, where $W$ is the bandwidth of the independently sampled disorder on each site~\cite{Abou-Chacra:1973qf}.
Altshuler, {\it et al.} argued that this result applies directly to the `hopping' problem in Fock space induced by two-body interactions in a disordered quantum dot~\cite{Altshuler:1997aa}.
They further conjectured a phase diagram in which the eigenstates are localized for $J < W/Z \log Z$, delocalized sparsely in Fock space for $W/Z\log Z < J < W/Z$ and fully delocalized for $W/Z < J$.

Physically, the transition at $W/Z\log Z$ arises due to long distance, high order resonances in the Fock space.
The associated states appear mixed from the point of view of few body observables, just as the GHZ state does, even though their participation ratios remain a vanishingly small fraction of the total volume of the hypercube.
They further would not exhibit level repulsion as states neighboring in energy typically do not overlap.

There are a number of assumptions that go into mapping the interacting dot model onto the Bethe lattice localization problem.
The most crucial of these are 1) that interference due to the presence of loops on the hypercube is unimportant at long distances and 2) that the energy correlations are likewise unimportant.
These are plausible but not rigorously justified.
While there is evidence from several detailed studies of related fully connected spin glass models for the $\log N$ enhancement of the critical coupling by higher order processes \cite{Burin:2016aa,Baldwin:2017aa,Laumann:2014aa}, the delocalized nonergodic phase remains controversial.
Indeed, even in the Bethe lattice problem, whether this regime is a phase or a slow crossover is still debated \cite{Biroli:2012vk,DeLuca:2014ch,Altshuler:2016aa}.

The central spin model can be mapped to the Bethe lattice problem under similar assumptions as the quantum dot model of Altshuler, {\it et al.} with $Z = N$.
This mapping predicts that higher order processes destabilize MBL as $N \to \infty$ at asymptotically smaller $J$ than the second order processes described in Sec.~\ref{sub:pt_second_order}.
If this is true, then it seems that there are two possibilities:
\begin{enumerate}
	\item There is a direct transition from the localized to the ETH phase at $J_c \sim W / N \log N$.  If this is the case, then it must be that at the small sizes $N$ that we can reach numerically, whatever non-perturbative processes cause the sparsely delocalized states to fully thermalize in the limit of large $N$ are ineffective at these small $N$, and we primarily observe behavior governed by the second-order picture up to the scale $J_c \sim W/N$.  This is why we develop a detailed picture of the second order percolation model in Sec.~\ref{sub:pt_consequences}.

	\item There is an intermediate delocalized ergodic phase between a transition at $J_c \sim W/ N \log N$ and another transition on the scale $J^* \sim W/N$. This would imply, for example, that $[S_1]$ transitions at asymptotically smaller coupling than $[r]$ by this factor of $\log N$.
\end{enumerate}

We have attempted to interpret our numerical results from both of these view points in Sec.~\ref{sec:numerics}, but the results are inconclusive, so we leave this difficult question about the possibility of such an intermediate phase in our central spin model for future work.

\section{Discussion}
\label{sec:discussion}

We have explored a central spin model for many-body localization and thermalization.  We show that a single central spin can serve as a `seed' for interactions between otherwise noninteracting l-bits and thus produce a thermal bath.  This bath, `nucleated' by just one central spin, can then thermalize an arbitrarily large number of otherwise noninteracting l-bits.  This is a limiting example of how a rare locally thermal region can destabilize a MBL phase \cite{De-Roeck:2017aa}.

We show, based both on numerics for small systems and on perturbative arguments, that a central-spin system with $N$ l-bits exhibits an MBL phase and phase transition in the limit of large $N$ if we scale the interaction down with $N$.  This dynamical transition occurs at an interaction strength that is thermodynamically insignificant in the large-$N$ limit.  We argue that the MBL phase can be understood as localization on a hypercube of Fock states \cite{Altshuler:1997aa}.  We discuss and explore the possibility of an intermediate ``delocalized but nonergodic'' phase between the MBL and thermal phases.  Our numerical results on small systems do not allow us to draw a conclusion about whether or not such an intermediate phase exists in this model.

We have also examined the finite-size behavior near the phase transition, particularly of the single-l-bit entanglement within eigenstates.  We find that the probability distribution of this quantity is bimodal near the transition, with peaks near full entanglement and near zero entanglement.  This heterogeneity is shown to occur both across l-bits within single eigenstates as well as across eigenstates for a single l-bit.  This is a feature of the finite-size data that has also been seen in 1D models and is not yet well understood \cite{Yu:2016aa,Khemani:2016aa}.  The finite-size data for our central spin model do not scale well, indicating that the sizes we can access numerically are well short of any asymptotic large-$N$ scaling regime.  One indication of this is that the sample-to-sample variations are small, but appear to be accelerating in their growth with increasing $N$.

Let us speculate on possible scenarios for the phase transition out of the localized phase in to a delocalized phase (which may be non-ergodic or obey ETH).
In the limit of large systems, each sample $s$ has its own transition point $J_c(s)$.
Due to the quenched disorder, we expect the probability distribution of these transition points to have a width $\Delta J_c \sim J_c/\sqrt{N}$ \cite{Chayes:1986kq,Chandran:2015aa}.
Within a single finite-size sample the transition may be sharper than this, of similar width, or broader than this.
The apparent accelerating increase of the sample-to-sample differences seen numerically (Fig.~\ref{fig:S1_bimodality}(c)) argues against the last of these possibilities.
This leaves the two other possibilities.
At the accessible system sizes, the inter-sample and within-sample variations are comparable; if this persists to large $N$, we would expect the rounding of the transition to be of width $\sim J_c/\sqrt{N}$ both within samples and between samples.
The trend with $N$ we see in Fig. \ref{fig:appendix_A_1}(b) suggests that the inhomogeneity between eigenstates in the l-bit-averaged entanglement at the transition might go away at large $N$, thus making all eigenstates in one sample similar in this respect (although it seems likely they will still differ in which l-bits are less vs. more entangled).

The other possible scenario is that the transition in a single sample is sharper by a power of $N$ than the variation in $J_c(s)$ between samples.
A limiting case of this scenario in equilibrium is provided by first-order transitions with quenched randomness \cite{Fisher:1995oq,Pazmandi:1997aa}.
In this scenario, if we scale $J$ by the sample-averaged $J_p$, then at $J_p$, almost all samples will be away from their transition and be in their MBL or delocalized phases respectively.
Almost all the variation in $S_1$ then comes from sample-to-sample variations.
We are clearly not in this regime for the sample sizes we study numerically here, but it remains possible that this happens at much larger $N$.
To study the single-sample transition in this regime, one should instead scale $J$ by the sample-specific transition point $J_c(s)$.
It is an interesting question to ask how sharp this transition could possibly be.  What would be the nature of the sharpest possible, thus most discontinuous, MBL-to-delocalized phase transition?  We leave these questions for future research.

\section{Acknowledgements}
We thank Wojciech de Roeck, Sarang Gopalakrishnan, Francois Huveneers and Vedika Khemani for helpful discussions.
 PP acknowledges financial support from Funda\c{c}\~ao para a
 Ci\^encia e a Tecnologia (Portugal) through grant SFRH/BD/84875/2012 and the Perimeter Institute for Theoretical Physics. Research at Perimeter
 Institute is supported through Industry Canada and by the Province of Ontario through the Ministry of Research \& Innovation.
 C.R.L. acknowledges support from the Sloan Foundation through a Sloan Research Fellowship and the NSF through grant PHY-1520535.
\appendix

\begin{widetext}

\section{Second-order processes}
\label{app:pt_second_order_details}

At second order in perturbation theory in $B$, there are $2 \binom{N}{2}$ configurations, $\ket{s' \hat{h}(\tau'_{ij}), \tau'_{ij}}$ connected to an initial state $\ket{s \hat{h}, \tau}$.
These correspond to flipping a pair of l-bits $i < j$ and, possibly flipping the central spin ($s'$) relative to its local field.
There are 4 channels contributing to the second order amplitude:
\newcommand{\ts}{\tilde{s}}
\begin{align}
\label{eq:secondorderamplitude}
\braket{s' \hat{h}(\tau'_{ij}), \tau'_{ij}}{s\hat{h}(\tau), \tau}^{(2)} &=
J^2 \sum_{\ts=\pm 1} \frac{\langle s' \hat{h}(\tau'_{ij})|B_i|\ts \hat{h}(\tau'_j)\rangle \langle \ts \hat{h}(\tau'_j)|B_j|s\hat{h}(\tau)\rangle}{(- \Delta E_{s' \tau'_{ij}})(- \Delta E_{\ts \tau'_j})} + (i \leftrightarrow j)
\end{align}
where $\Delta E_{s'\tau'_j} = E_{s'\tau'_j} - E_{s\tau}$ are the changes in energy relative to the initial state.

Naively, the final energy denominator, $\Delta E_{s' \tau'_{ij}}$, takes $\sim 2N^2$ values on a band of width $W$ and thus has density of states $\sim 2N^2/W$.
The intermediate energy differences, $\Delta E_{\ts, \tau'}$, are typically $O(W)$ and the numerators are $O(J^2)$ so large amplitudes (resonances) arise in Eq.~\eqref{eq:secondorderamplitude} for $\sim J^2 N^2/ W^2$ final states.
This predicts $O(1)$ second order resonances when
\begin{align}
	J \sim W/N
\end{align}

It turns out that this argument is essentially correct for those final states in which the central spin is flipped $s' = -s$.
However, destructive interference between the four channels reduces the probability of resonance for final states in which the central spin is not flipped $s' = s$.
Indeed, such cancellations must prevent second order resonance if the two l-bits involved were non-interacting and thus in product states.
Since the interaction terms in this model are $O(J)$, care must be taken to check the interference for $J \sim W/N$.

More precisely, consider the interference between two second-order channels with intermediate states $1$ and $2$, respectively, to reach a final state $12$,
\begin{align}
	\frac{\mathcal{M}(1+\eta)}{(-\Delta E_{12})(-\Delta E_1)}+\frac{\mathcal{M}(1-\eta)}{(-\Delta E_{12})(-\Delta E_2)}
\end{align}
where $\Delta E_{12} = \Delta E_1 + \Delta E_2 + \Delta^{(12)}$ and $\mathcal{M}(1\pm\eta)$ are the numerators of the respective processes. The amplitude can be re-written as:
\begin{align}
\label{eq:resonancecancel}
	\frac{\mathcal{M}}{\Delta E_1 \Delta E_2}\left(1 - \frac{\Delta^{(12)}}{\Delta E_{12}} + \eta \frac{\Delta E_2 - \Delta E_1}{\Delta E_{12}}\right)
\end{align}
The first term does not involve the final energy denominator $\Delta E_{12}$ at all and thus does not cause second order resonance (assuming that the first order differences $\Delta E_1, \Delta E_2$ are themselves not resonant).
Resonances only arise due to interaction energy $\Delta^{(12)}$ and matrix element deviations $\eta$. If both of these are parametrically small in $N$, then resonances are suppressed.

Let us apply Eq.~\eqref{eq:resonancecancel} to the determination of resonances in Eq.~\eqref{eq:secondorderamplitude}. There are two types of final states depending on whether the central spin is flipped ($s' = - s$) or not ($s' = s$) with respect to the initial configuration.

\paragraph{Central spin unflipped ($s' = s$)---}%
In this case, the four channels destructively interfere in pairs according to the intermediate central spin state $\ts$.

For the pair of channels with $\ts = s$,
\begin{align}
	\Delta E_{s \tau'_{ij}} = \Delta E_{s \tau'_i} + \Delta E_{s \tau'_j} + s \frac{J^2}{|\vec{h}|}( \vec{\delta}_i \cdot \vec{\delta}_j - (\hat{h}\cdot \vec{\delta}_i)(\hat{h}\cdot \vec{\delta}_j)) + \cdots
\end{align}
so that
\begin{align}
	\Delta^{(ij)} = s \frac{J^2}{|\vec{h}|}( \vec{\delta}_i \cdot \vec{\delta}_j - (\hat{h}\cdot \vec{\delta}_i)(\hat{h}\cdot \vec{\delta}_j)) + \cdots  = O\left(\frac{J}{\sqrt{N}}\right) \label{Eq:IntEgyUnflippedInt}
\end{align}
To leading order in $1/N$, the matrix elements
\begin{align}
	\mathcal{M} &= J^2 B_i^{++} B_j^{++} + O(J^2/\sqrt{N}) \\
	\eta &= O(1/\sqrt{N})
\end{align}
where we have used that $\ket{\hat{h}(\tau')} = \ket{\hat{h}(\tau)} + O(1/\sqrt{N})$ for $\tau' = \tau'_i, \tau'_j, \tau'_{ij}$. We have also introduced the short hand notation for the matrix elements of $B_i$ with respect to the initial central spin field $\hat{h}(\tau)$,
\begin{align}
	B_i^{\pm \pm} = \bra{\pm s \hat{h}} B_i \ket{\pm s \hat{h}}
\end{align}

For the pair of channels with $\ts = -s$,
\begin{align}
	\Delta E_{s \tau'_{ij}} = \Delta E_{-s \tau'_i} + \Delta E_{-s \tau'_j} + 4 s |\vec{h}| + \cdots
\end{align}
which provides a larger interaction energy as compared to Eq.~\eqref{Eq:IntEgyUnflippedInt}:
\begin{align}
	\Delta^{(ij)} &= 4 s |\vec{h}| + \cdots = O(J \sqrt{N})
\end{align}
The matrix elements are:
\begin{align}
	\mathcal{M} &= \frac{J^2}{2} \left(B_i^{+-} B_j^{-+} + B_j^{+-} B_i^{-+}\right)+ O(J^2/\sqrt{N}) \\
	\eta &= \frac{B_i^{+-} B_j^{-+} - B_j^{+-} B_i^{-+}}{B_i^{+-} B_j^{-+} + B_j^{+-} B_i^{-+}} + O(1/\sqrt{N})
\end{align}
For the case of GOE random matrices $B$ treated in this manuscript, this simplifies to $\eta = O(1/\sqrt{N})$ as $B_{i,j}^{+-} = B_{i,j}^{-+}$.

With reference to Eq.~\eqref{eq:resonancecancel}, the largest amplitudes arise from the $\Delta^{(12)}$ interaction term for the $\ts = -s$ pair of channels.
Its contribution to the amplitude exceeds $1$ when
\begin{align}
	\frac{J^2}{W^2}\times \frac{J \sqrt{N}}{W/N^2} &\gtrsim 1 \\
	\Rightarrow J &\gtrsim W/N^{5/6}
\end{align}
Thus, at $J \sim W/N \ll W/N^{5/6}$, there are no second order resonances for final states with the central spin unflipped.

\paragraph{Central spin flipped ($s'=-s$)---}
In this case, rather than being paired by the intermediate state of the central spin $\ts$, it is natural to pair the channels by whether the central spin flips simultaneously with the $i$'th l-bit or with the $j$'th.
This is because these pairings lead to the smallest interaction energy $\Delta^{(ij)}$ and correspondingly most destructive interference:
\begin{align}
	\Delta E_{-s \tau''_{ij}} &= \Delta E_{-s \tau'_i} + \Delta E_{s \tau'_j} + O(J) \\
	\Delta E_{-s \tau''_{ij}} &= \Delta E_{s \tau'_i} + \Delta E_{-s \tau'_j} + O(J)
\end{align}
The interaction energy is only $\Delta^{(ij)} = O(J)$, which is again small for $J \sim W/N$.

However, unlike the previous cases, the matrix elements for the four channels are quite different. For example, in the first pair (central spin flips with $i$):
\begin{align}
	\mathcal{M} &= \frac{J^2}{2} \left(B_i^{-+}B_j^{++} + B_j^{--}B_i^{-+}\right) + \cdots \\
	\eta &= \frac{B_i^{-+}B_j^{++} - B_j^{--}B_i^{-+}}{B_i^{-+}B_j^{++} + B_j^{--}B_i^{-+}} + \cdots = O(1)
\end{align}
where the last step follows because $B_j^{--}$ and $B_j^{++}$ are independent Gaussian random numbers of $O(1)$.

This suggests that the amplitude in Eq.~\eqref{eq:resonancecancel} can become large because of the $\eta$ term already if
\begin{align}
	\frac{J^2}{W^2} \times O(1) \times \frac{W}{W/N^2} \gtrsim 1
\end{align}
That is, for
\begin{align}
	J \gtrsim W/N
\end{align}
as expected by the naive argument.

Let us check that no further cancellations arise from summing all four channels.
For simplicity, we may assume that the interaction energies $\Delta^{(ij)}$ are zero, as this only increases the destructive interference. In this case, we have that $\Delta E_{s \tau'_i} \approx -\Delta E_{-s \tau'_j}$ and $\Delta E_{-s \tau'_i} \approx -\Delta E_{s \tau'_j}$. Thus, the sum of the two most divergent ($\eta$) terms in Eq.~\eqref{eq:resonancecancel} is (to leading order)
\begin{align}
	 \frac{J^2}{\Delta E_{-s \tau'_{ij}}} \left[ \frac{B_i^{-+}B_j^{++}-B_j^{--}B_i^{-+}}{\Delta E_{-s \tau'_i}} +
	\frac{B_j^{-+}B_i^{++}-B_i^{--}B_j^{-+}}{\Delta E_{s \tau'_i}} \right] + \cdots
\end{align}
The two terms in the square brackets are independent random variables of $O(1)$. Thus, they typically add in quadrature and there is no cancellation.
This term is therefore $\sim \frac{J^2}{W/N^2} \times \frac{1}{W}$, which reproduces the estimate that $J \sim W/N$ produces $O(1)$ resonances.

\section{Properties of single l-bit entanglement entropy}
\label{app:std_dev_parsing}

In this appendix, we parse the contributions to the standard deviation $\Delta S_1$ of the single l-bit entanglement entropy into the components arising due to differences between l-bits, between eigenstates, and between samples \cite{Khemani:2016aa}.

We find that for our range of $N$ the largest contributions to $\Delta S_1$ near the transition come from variations between l-bits in the same eigenstate, i.e. $[\Delta_l S_1]_{E,s}$ (Fig. \ref{fig:appendix_A_1}(a)), and across different eigenstates for the same l-bit, i.e. $[\Delta_E S_1^i]_{s}$ (Fig.~\ref{fig:S1_bimodality}(b)).
This is in contrast with previously studied 1D models with quenched disorder exhibiting an MBL-thermal transition: in Ref.~\cite{Khemani:2016aa}, the sample-to-sample variation is the dominant contribution at the largest sizes accessible to diagonalization.

The eigenstate dependence of the l-bit entropies $S_1$, captured by $[\Delta_{E}[S_1]_{l}]_s$ also peaks at the transition (Fig. \ref{fig:appendix_A_1}(b)). However, it is smaller than $[\Delta_l S_1]_{E,s}$ and slightly decreases with $N$, suggesting that the variation between eigenstates in their overall degree of thermalization is a smaller contributor to $\Delta S_1$ as compared to within-eigenstate variations between l-bits.

For completeness and ease of comparison, we show again as Fig. \ref{fig:appendix_A_1}(c) the sample-to-sample variations which were shown already in the main text in Fig.~\ref{fig:S1_bimodality}(c).

\begin{figure*}[h]
\includegraphics[width=0.329\textwidth]{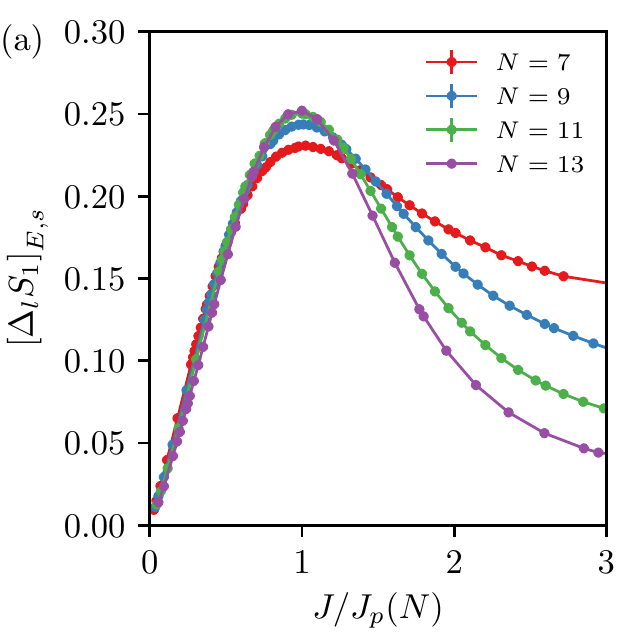}
\includegraphics[width=0.329\textwidth]{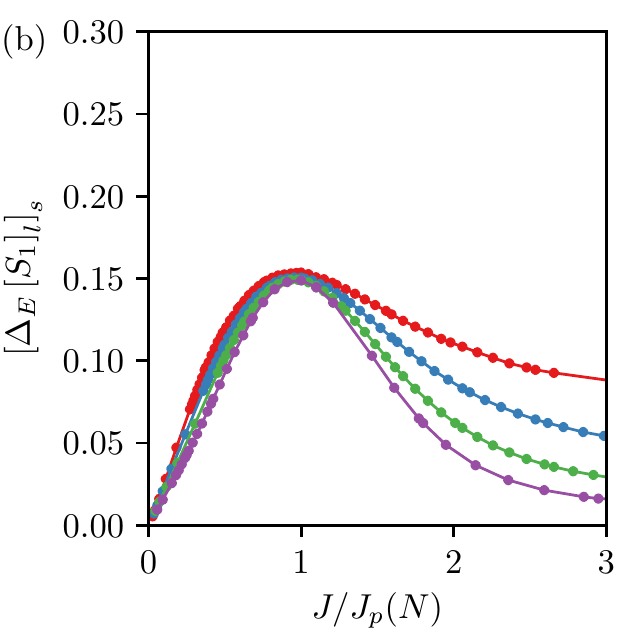}
\includegraphics[width=0.329\textwidth]{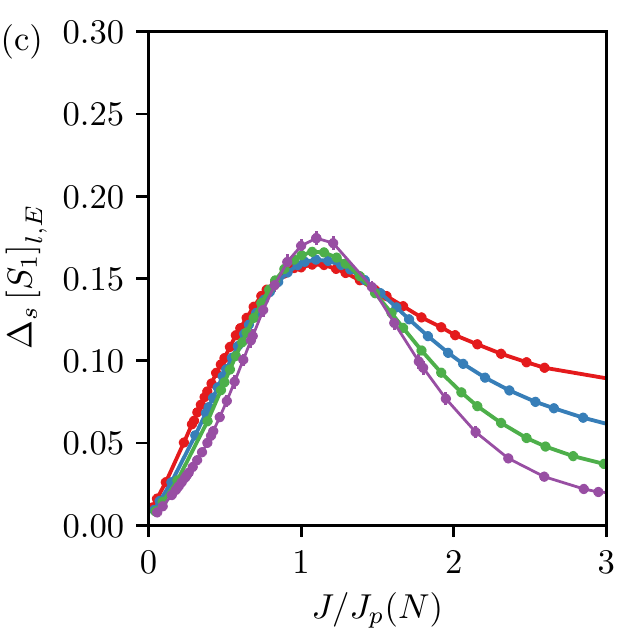}
\caption{\label{fig:appendix_A_1} (a) Standard deviation of $S_1$ over l-bits in the same eigenstate, averaged over different eigenstates and samples. (b) Average over samples of the standard deviation over eigenstates of the l-bit-averaged $S_1$.   (c) Standard deviation over samples of $S_1$ averaged within each sample over eigenstates and l-bits.}
\end{figure*}

\section{Central spin entanglement}
\label{app:central_spin_entanglement}

The focus of this appendix is the behavior of the central spin entanglement in eigenstates $S_{\mathrm{cs}}$.
At $B_i=0$, the eigenstates $|s\hat{h}(\tau), \tau\rangle$ are product states between the central spin and the l-bits with $S_{\mathrm{cs}}=0$ in all eigenstates and samples.
At small $B_i$, the central spin flips in the resonant processes that hybridize the classical states (see Sec.~\ref{sec:perturbative_analysis}).
As the central spin is substantially entangled in eigenstates that are spread over a few classical states, we expect that $S_{\mathrm{cs}}$ is close to one in eigenstates with even a few resonant l-bits, while it is close to zero in eigenstates with no resonant l-bits.
This has two important consequences:
(1) the moments of the distribution of $S_{\mathrm{cs}}$ are non-zero in the MBL phase even in the thermodynamic limit, and
(2) the central spin crosses over to high entanglement at $J^{\mathrm{cs}}_p(N) < J_p(N)$ within the MBL phase.
Note however that the scaling of $J^{\mathrm{cs}}_p(N)$ with $N$ is the same as that of $J_p(N)$; that is, $J^{\mathrm{cs}}_p(N) \sim W/N$.

Feature (1) is already visible in the mean $[S_{\mathrm{cs}}]$ of the central spin entanglement (see Fig.~\ref{fig:appendix_B_1}(a)) where the $N$ dependence only emerges for $J \sim J_p(N)$.
The total standard deviation $\Delta S_{\mathrm{cs}}$ brings out feature (2) more clearly (see Fig.~\ref{fig:appendix_B_1}(b)) as the peak lies at $J \sim J^{\mathrm{cs}}_p(N) < J_p(N)$ to the left of the global transition point.
Furthermore, the peak value of $\Delta S_{\mathrm{cs}}$ exceeds $0.28$ (the standard deviation of a uniform distribution), showing that the distribution is bi-modal in the vicinity of $J^{\mathrm{cs}}_p(N)$.
Finally, in the delocalized phase at $J/J_p(N)>1$, $\Delta S_{\mathrm{cs}}$ decreases with increasing $N$, in agreement with ETH.

In order to bring out the sharp contrast between $\Delta S_1$ of the l-bits (Fig.~\ref{fig:Rescaled_DeltaS_Peak}) and  $\Delta S_{\mathrm{cs}}$ of the central spin, we plot $\Delta S_{\mathrm{cs}}$ normalized by its peak value vs $J/J^{\mathrm{cs}}_{p}(N)$ in Fig. \ref{fig:appendix_B_1}(b).
Unlike the l-bit entanglement entropy variation, $\Delta S_{\mathrm{cs}}$ does not narrow on its own scale with increasing $N$ in the MBL phase.
Hence the central spin's entanglement undergoes a crossover within the MBL phase, as opposed to the l-bits for which the peaks narrow as $N$ is increased as seen in Fig.~\ref{fig:Rescaled_DeltaS_Peak}.

\begin{figure*}[h]
\includegraphics[width=0.329\textwidth]{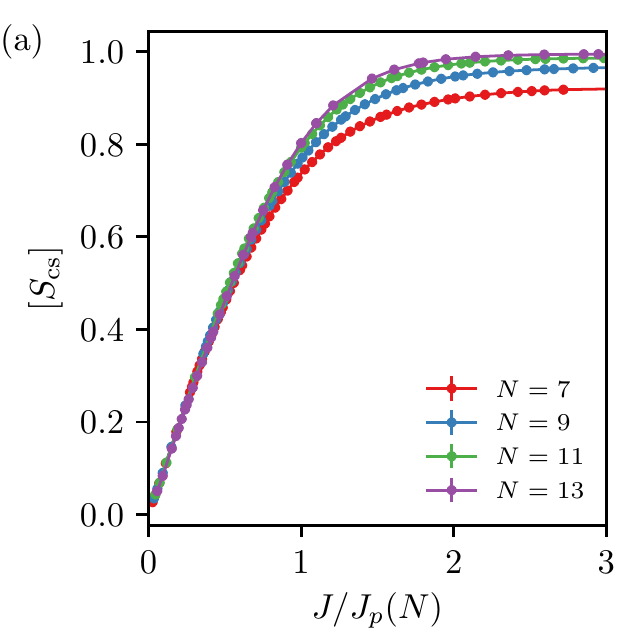}
\includegraphics[width=0.329\textwidth]{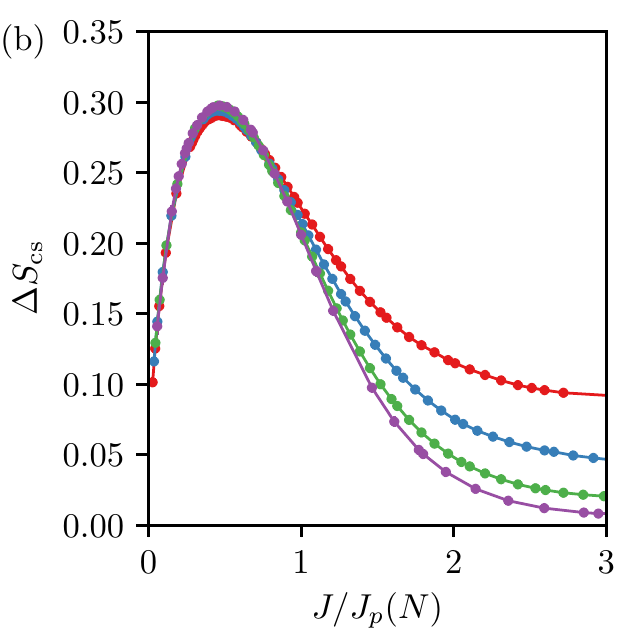}
\includegraphics[width=0.329\textwidth]{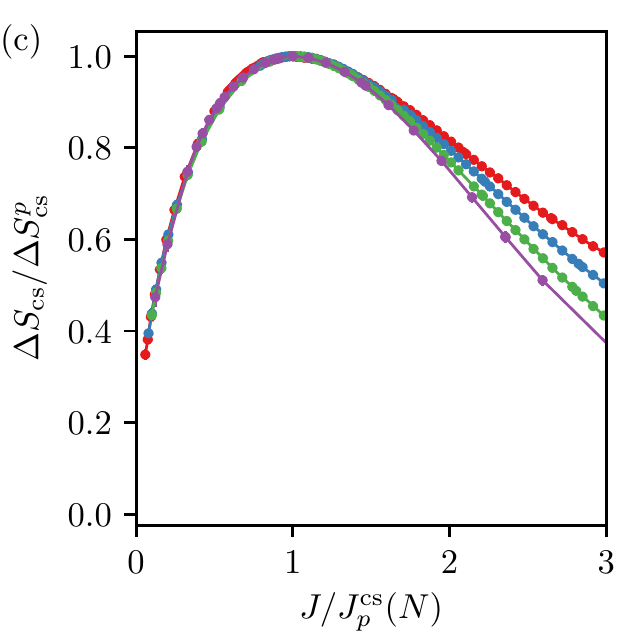}
\caption{\label{fig:appendix_B_1} (a) Mean entanglement entropy of the central spin $[S_{\mathrm{cs}}]$ and
(b) standard deviation  $\Delta S_{\mathrm{cs}}$  vs. scaled coupling $J/J_p(N)$.
(c) $\Delta S_{\mathrm{cs}}$ normalized by its value at the peak for a given number of l-bits $N$ vs the coupling $J$ on the scale of the coupling $J^{\mathrm{cs}}_{p}(N)$.
}

\end{figure*}

\end{widetext}

%%%%%%%%%%%%%%%%%%%%%
\bibliography{papers}

\begin{thebibliography}{66}
\expandafter\ifx\csname natexlab\endcsname\relax\def\natexlab#1{#1}\fi
\expandafter\ifx\csname bibnamefont\endcsname\relax
  \def\bibnamefont#1{#1}\fi
\expandafter\ifx\csname bibfnamefont\endcsname\relax
  \def\bibfnamefont#1{#1}\fi
\expandafter\ifx\csname citenamefont\endcsname\relax
  \def\citenamefont#1{#1}\fi
\expandafter\ifx\csname url\endcsname\relax
  \def\url#1{\texttt{#1}}\fi
\expandafter\ifx\csname urlprefix\endcsname\relax\def\urlprefix{URL }\fi
\providecommand{\bibinfo}[2]{#2}
\providecommand{\eprint}[2][]{\url{#2}}

\bibitem[{\citenamefont{Anderson}(1958)}]{Anderson:1958ly}
\bibinfo{author}{\bibfnamefont{P.~W.} \bibnamefont{Anderson}},
  \bibinfo{journal}{Phys. Rev.} \textbf{\bibinfo{volume}{109}},
  \bibinfo{pages}{1492} (\bibinfo{year}{1958}).

\bibitem[{\citenamefont{Gornyi et~al.}(2005)\citenamefont{Gornyi, Mirlin, and
  Polyakov}}]{Gornyi:2005lq}
\bibinfo{author}{\bibfnamefont{I.~V.} \bibnamefont{Gornyi}},
  \bibinfo{author}{\bibfnamefont{A.~D.} \bibnamefont{Mirlin}},
  \bibnamefont{and} \bibinfo{author}{\bibfnamefont{D.~G.}
  \bibnamefont{Polyakov}}, \bibinfo{journal}{Phys. Rev. Lett.}
  \textbf{\bibinfo{volume}{95}}, \bibinfo{pages}{206603}
  (\bibinfo{year}{2005}).

\bibitem[{\citenamefont{Basko et~al.}(2006)\citenamefont{Basko, Aleiner, and
  Altshuler}}]{Basko:2006aa}
\bibinfo{author}{\bibfnamefont{D.}~\bibnamefont{Basko}},
  \bibinfo{author}{\bibfnamefont{I.}~\bibnamefont{Aleiner}}, \bibnamefont{and}
  \bibinfo{author}{\bibfnamefont{B.}~\bibnamefont{Altshuler}},
  \bibinfo{journal}{Annals of Physics} \textbf{\bibinfo{volume}{321}},
  \bibinfo{pages}{1126 } (\bibinfo{year}{2006}), ISSN
  \bibinfo{issn}{0003-4916}.

\bibitem[{\citenamefont{Oganesyan and Huse}(2007)}]{Oganesyan:2007aa}
\bibinfo{author}{\bibfnamefont{V.}~\bibnamefont{Oganesyan}} \bibnamefont{and}
  \bibinfo{author}{\bibfnamefont{D.~A.} \bibnamefont{Huse}},
  \bibinfo{journal}{Phys. Rev. B} \textbf{\bibinfo{volume}{75}},
  \bibinfo{pages}{155111} (\bibinfo{year}{2007}).

\bibitem[{\citenamefont{Pal and Huse}(2010)}]{Pal:2010gs}
\bibinfo{author}{\bibfnamefont{A.}~\bibnamefont{Pal}} \bibnamefont{and}
  \bibinfo{author}{\bibfnamefont{D.~A.} \bibnamefont{Huse}},
  \bibinfo{journal}{Phys. Rev. B} \textbf{\bibinfo{volume}{82}},
  \bibinfo{pages}{174411} (\bibinfo{year}{2010}).

\bibitem[{\citenamefont{Nandkishore and Huse}(2015)}]{Nandkishore:2015aa}
\bibinfo{author}{\bibfnamefont{R.}~\bibnamefont{Nandkishore}} \bibnamefont{and}
  \bibinfo{author}{\bibfnamefont{D.~A.} \bibnamefont{Huse}},
  \bibinfo{journal}{Annual Review of Condensed Matter Physics}
  \textbf{\bibinfo{volume}{6}}, \bibinfo{pages}{15} (\bibinfo{year}{2015}),
  ISSN \bibinfo{issn}{1947-5462}.

\bibitem[{\citenamefont{Altman and Vosk}(2015)}]{Altman:2015aa}
\bibinfo{author}{\bibfnamefont{E.}~\bibnamefont{Altman}} \bibnamefont{and}
  \bibinfo{author}{\bibfnamefont{R.}~\bibnamefont{Vosk}},
  \bibinfo{journal}{Annual Review of Condensed Matter Physics}
  \textbf{\bibinfo{volume}{6}}, \bibinfo{pages}{383} (\bibinfo{year}{2015}),
  ISSN \bibinfo{issn}{1947-5462}.

\bibitem[{\citenamefont{Serbyn et~al.}(2013{\natexlab{a}})\citenamefont{Serbyn,
  Papi\ifmmode~\acute{c}\else \'{c}\fi{}, and Abanin}}]{Serbyn:2013rt}
\bibinfo{author}{\bibfnamefont{M.}~\bibnamefont{Serbyn}},
  \bibinfo{author}{\bibfnamefont{Z.}~\bibnamefont{Papi\ifmmode~\acute{c}\else
  \'{c}\fi{}}}, \bibnamefont{and} \bibinfo{author}{\bibfnamefont{D.~A.}
  \bibnamefont{Abanin}}, \bibinfo{journal}{Phys. Rev. Lett.}
  \textbf{\bibinfo{volume}{111}}, \bibinfo{pages}{127201}
  (\bibinfo{year}{2013}{\natexlab{a}}).

\bibitem[{\citenamefont{Huse et~al.}(2014)\citenamefont{Huse, Nandkishore, and
  Oganesyan}}]{Huse:2014ac}
\bibinfo{author}{\bibfnamefont{D.~A.} \bibnamefont{Huse}},
  \bibinfo{author}{\bibfnamefont{R.}~\bibnamefont{Nandkishore}},
  \bibnamefont{and}
  \bibinfo{author}{\bibfnamefont{V.}~\bibnamefont{Oganesyan}},
  \bibinfo{journal}{Physical Review B} \textbf{\bibinfo{volume}{90}}
  (\bibinfo{year}{2014}), ISSN \bibinfo{issn}{1550-235X}.

\bibitem[{\citenamefont{Imbrie}(2016)}]{Imbrie2016}
\bibinfo{author}{\bibfnamefont{J.~Z.} \bibnamefont{Imbrie}},
  \bibinfo{journal}{Journal of Statistical Physics}
  \textbf{\bibinfo{volume}{163}}, \bibinfo{pages}{998} (\bibinfo{year}{2016}),
  ISSN \bibinfo{issn}{1572-9613}.

\bibitem[{\citenamefont{Ros et~al.}(2015)\citenamefont{Ros, Muller, and
  Scardicchio}}]{Ros:2015rw}
\bibinfo{author}{\bibfnamefont{V.}~\bibnamefont{Ros}},
  \bibinfo{author}{\bibfnamefont{M.}~\bibnamefont{Muller}}, \bibnamefont{and}
  \bibinfo{author}{\bibfnamefont{A.}~\bibnamefont{Scardicchio}},
  \bibinfo{journal}{Nuclear Physics B} \textbf{\bibinfo{volume}{891}},
  \bibinfo{pages}{420 } (\bibinfo{year}{2015}), ISSN \bibinfo{issn}{0550-3213}.

\bibitem[{\citenamefont{Chandran et~al.}(2015)\citenamefont{Chandran, Kim,
  Vidal, and Abanin}}]{chandran2015constructing}
\bibinfo{author}{\bibfnamefont{A.}~\bibnamefont{Chandran}},
  \bibinfo{author}{\bibfnamefont{I.~H.} \bibnamefont{Kim}},
  \bibinfo{author}{\bibfnamefont{G.}~\bibnamefont{Vidal}}, \bibnamefont{and}
  \bibinfo{author}{\bibfnamefont{D.~A.} \bibnamefont{Abanin}},
  \bibinfo{journal}{Physical Review B} \textbf{\bibinfo{volume}{91}},
  \bibinfo{pages}{085425} (\bibinfo{year}{2015}).

\bibitem[{\citenamefont{Monthus}(2016)}]{Monthus:2016aa}
\bibinfo{author}{\bibfnamefont{C.}~\bibnamefont{Monthus}},
  \bibinfo{journal}{Journal of Statistical Mechanics: Theory and Experiment}
  \textbf{\bibinfo{volume}{2016}}, \bibinfo{pages}{033101}
  (\bibinfo{year}{2016}).

\bibitem[{\citenamefont{Rademaker and Ortu\~no}(2016)}]{Rademaker:2016aa}
\bibinfo{author}{\bibfnamefont{L.}~\bibnamefont{Rademaker}} \bibnamefont{and}
  \bibinfo{author}{\bibfnamefont{M.}~\bibnamefont{Ortu\~no}},
  \bibinfo{journal}{Phys. Rev. Lett.} \textbf{\bibinfo{volume}{116}},
  \bibinfo{pages}{010404} (\bibinfo{year}{2016}).

\bibitem[{\citenamefont{Znidaric et~al.}(2008)\citenamefont{Znidaric, Prosen,
  and Prelovsek}}]{Znidaric:2008aa}
\bibinfo{author}{\bibfnamefont{M.}~\bibnamefont{Znidaric}},
  \bibinfo{author}{\bibfnamefont{T.}~\bibnamefont{Prosen}}, \bibnamefont{and}
  \bibinfo{author}{\bibfnamefont{P.}~\bibnamefont{Prelovsek}},
  \bibinfo{journal}{Phys. Rev. B} \textbf{\bibinfo{volume}{77}},
  \bibinfo{pages}{064426} (\bibinfo{year}{2008}).

\bibitem[{\citenamefont{Bardarson et~al.}(2012)\citenamefont{Bardarson,
  Pollmann, and Moore}}]{Bardarson:2012kl}
\bibinfo{author}{\bibfnamefont{J.~H.} \bibnamefont{Bardarson}},
  \bibinfo{author}{\bibfnamefont{F.}~\bibnamefont{Pollmann}}, \bibnamefont{and}
  \bibinfo{author}{\bibfnamefont{J.~E.} \bibnamefont{Moore}},
  \bibinfo{journal}{Phys. Rev. Lett.} \textbf{\bibinfo{volume}{109}},
  \bibinfo{pages}{017202} (\bibinfo{year}{2012}).

\bibitem[{\citenamefont{Bauer and Nayak}(2013)}]{Bauer:2013rz}
\bibinfo{author}{\bibfnamefont{B.}~\bibnamefont{Bauer}} \bibnamefont{and}
  \bibinfo{author}{\bibfnamefont{C.}~\bibnamefont{Nayak}},
  \bibinfo{journal}{Journal of Statistical Mechanics: Theory and Experiment}
  \textbf{\bibinfo{volume}{2013}}, \bibinfo{pages}{P09005}
  (\bibinfo{year}{2013}).

\bibitem[{\citenamefont{Serbyn et~al.}(2013{\natexlab{b}})\citenamefont{Serbyn,
  Papi\ifmmode~\acute{c}\else \'{c}\fi{}, and Abanin}}]{Serbyn:2013uq}
\bibinfo{author}{\bibfnamefont{M.}~\bibnamefont{Serbyn}},
  \bibinfo{author}{\bibfnamefont{Z.}~\bibnamefont{Papi\ifmmode~\acute{c}\else
  \'{c}\fi{}}}, \bibnamefont{and} \bibinfo{author}{\bibfnamefont{D.~A.}
  \bibnamefont{Abanin}}, \bibinfo{journal}{Phys. Rev. Lett.}
  \textbf{\bibinfo{volume}{110}}, \bibinfo{pages}{260601}
  (\bibinfo{year}{2013}{\natexlab{b}}).

\bibitem[{\citenamefont{Huse et~al.}(2013)\citenamefont{Huse, Nandkishore,
  Oganesyan, Pal, and Sondhi}}]{Huse:2013aa}
\bibinfo{author}{\bibfnamefont{D.~A.} \bibnamefont{Huse}},
  \bibinfo{author}{\bibfnamefont{R.}~\bibnamefont{Nandkishore}},
  \bibinfo{author}{\bibfnamefont{V.}~\bibnamefont{Oganesyan}},
  \bibinfo{author}{\bibfnamefont{A.}~\bibnamefont{Pal}}, \bibnamefont{and}
  \bibinfo{author}{\bibfnamefont{S.~L.} \bibnamefont{Sondhi}},
  \bibinfo{journal}{Phys. Rev. B} \textbf{\bibinfo{volume}{88}},
  \bibinfo{pages}{014206} (\bibinfo{year}{2013}).

\bibitem[{\citenamefont{Pekker et~al.}(2014)\citenamefont{Pekker, Refael,
  Altman, Demler, and Oganesyan}}]{Pekker:2014aa}
\bibinfo{author}{\bibfnamefont{D.}~\bibnamefont{Pekker}},
  \bibinfo{author}{\bibfnamefont{G.}~\bibnamefont{Refael}},
  \bibinfo{author}{\bibfnamefont{E.}~\bibnamefont{Altman}},
  \bibinfo{author}{\bibfnamefont{E.}~\bibnamefont{Demler}}, \bibnamefont{and}
  \bibinfo{author}{\bibfnamefont{V.}~\bibnamefont{Oganesyan}},
  \bibinfo{journal}{Phys. Rev. X} \textbf{\bibinfo{volume}{4}},
  \bibinfo{pages}{011052} (\bibinfo{year}{2014}).

\bibitem[{\citenamefont{Chandran et~al.}(2014)\citenamefont{Chandran, Khemani,
  Laumann, and Sondhi}}]{Chandran:2014aa}
\bibinfo{author}{\bibfnamefont{A.}~\bibnamefont{Chandran}},
  \bibinfo{author}{\bibfnamefont{V.}~\bibnamefont{Khemani}},
  \bibinfo{author}{\bibfnamefont{C.~R.} \bibnamefont{Laumann}},
  \bibnamefont{and} \bibinfo{author}{\bibfnamefont{S.~L.}
  \bibnamefont{Sondhi}}, \bibinfo{journal}{Phys. Rev. B}
  \textbf{\bibinfo{volume}{89}}, \bibinfo{pages}{144201}
  (\bibinfo{year}{2014}).

\bibitem[{\citenamefont{Kj\"all et~al.}(2014)\citenamefont{Kj\"all, Bardarson,
  and Pollmann}}]{Kjall:2014aa}
\bibinfo{author}{\bibfnamefont{J.~A.} \bibnamefont{Kj\"all}},
  \bibinfo{author}{\bibfnamefont{J.~H.} \bibnamefont{Bardarson}},
  \bibnamefont{and} \bibinfo{author}{\bibfnamefont{F.}~\bibnamefont{Pollmann}},
  \bibinfo{journal}{Phys. Rev. Lett.} \textbf{\bibinfo{volume}{113}},
  \bibinfo{pages}{107204} (\bibinfo{year}{2014}).

\bibitem[{\citenamefont{Bahri et~al.}(2015)\citenamefont{Bahri, Vosk, Altman,
  and Vishwanath}}]{Bahri:2015aa}
\bibinfo{author}{\bibfnamefont{Y.}~\bibnamefont{Bahri}},
  \bibinfo{author}{\bibfnamefont{R.}~\bibnamefont{Vosk}},
  \bibinfo{author}{\bibfnamefont{E.}~\bibnamefont{Altman}}, \bibnamefont{and}
  \bibinfo{author}{\bibfnamefont{A.}~\bibnamefont{Vishwanath}},
  \bibinfo{journal}{Nat Commun} \textbf{\bibinfo{volume}{6}}
  (\bibinfo{year}{2015}).

\bibitem[{\citenamefont{Bar~Lev et~al.}(2015)\citenamefont{Bar~Lev, Cohen, and
  Reichman}}]{Bar-Lev:2015aa}
\bibinfo{author}{\bibfnamefont{Y.}~\bibnamefont{Bar~Lev}},
  \bibinfo{author}{\bibfnamefont{G.}~\bibnamefont{Cohen}}, \bibnamefont{and}
  \bibinfo{author}{\bibfnamefont{D.~R.} \bibnamefont{Reichman}},
  \bibinfo{journal}{Phys. Rev. Lett.} \textbf{\bibinfo{volume}{114}},
  \bibinfo{pages}{100601} (\bibinfo{year}{2015}).

\bibitem[{\citenamefont{Devakul and Singh}(2015)}]{Devakul:2015aa}
\bibinfo{author}{\bibfnamefont{T.}~\bibnamefont{Devakul}} \bibnamefont{and}
  \bibinfo{author}{\bibfnamefont{R.~R.~P.} \bibnamefont{Singh}},
  \bibinfo{journal}{Phys. Rev. Lett.} \textbf{\bibinfo{volume}{115}},
  \bibinfo{pages}{187201} (\bibinfo{year}{2015}).

\bibitem[{\citenamefont{Torres-Herrera and
  Santos}(2015)}]{Torres-Herrera:2015aa}
\bibinfo{author}{\bibfnamefont{E.~J.} \bibnamefont{Torres-Herrera}}
  \bibnamefont{and} \bibinfo{author}{\bibfnamefont{L.~F.}
  \bibnamefont{Santos}}, \bibinfo{journal}{Phys. Rev. B}
  \textbf{\bibinfo{volume}{92}}, \bibinfo{pages}{014208}
  (\bibinfo{year}{2015}).

\bibitem[{\citenamefont{Gopalakrishnan
  et~al.}(2016)\citenamefont{Gopalakrishnan, Agarwal, Demler, Huse, and
  Knap}}]{Gopalakrishnan:2016aa}
\bibinfo{author}{\bibfnamefont{S.}~\bibnamefont{Gopalakrishnan}},
  \bibinfo{author}{\bibfnamefont{K.}~\bibnamefont{Agarwal}},
  \bibinfo{author}{\bibfnamefont{E.~A.} \bibnamefont{Demler}},
  \bibinfo{author}{\bibfnamefont{D.~A.} \bibnamefont{Huse}}, \bibnamefont{and}
  \bibinfo{author}{\bibfnamefont{M.}~\bibnamefont{Knap}},
  \bibinfo{journal}{Phys. Rev. B} \textbf{\bibinfo{volume}{93}},
  \bibinfo{pages}{134206} (\bibinfo{year}{2016}).

\bibitem[{\citenamefont{Luitz et~al.}(2016)\citenamefont{Luitz, Laflorencie,
  and Alet}}]{Luitz:2016aa}
\bibinfo{author}{\bibfnamefont{D.~J.} \bibnamefont{Luitz}},
  \bibinfo{author}{\bibfnamefont{N.}~\bibnamefont{Laflorencie}},
  \bibnamefont{and} \bibinfo{author}{\bibfnamefont{F.}~\bibnamefont{Alet}},
  \bibinfo{journal}{Phys. Rev. B} \textbf{\bibinfo{volume}{93}},
  \bibinfo{pages}{060201} (\bibinfo{year}{2016}).

\bibitem[{\citenamefont{Znidaric et~al.}(2016)\citenamefont{Znidaric,
  Scardicchio, and Varma}}]{Znidaric:2016aa}
\bibinfo{author}{\bibfnamefont{M.}~\bibnamefont{Znidaric}},
  \bibinfo{author}{\bibfnamefont{A.}~\bibnamefont{Scardicchio}},
  \bibnamefont{and} \bibinfo{author}{\bibfnamefont{V.~K.} \bibnamefont{Varma}},
  \bibinfo{journal}{Phys. Rev. Lett.} \textbf{\bibinfo{volume}{117}},
  \bibinfo{pages}{040601} (\bibinfo{year}{2016}).

\bibitem[{\citenamefont{Altland and Micklitz}(2017)}]{Altland:2017aa}
\bibinfo{author}{\bibfnamefont{A.}~\bibnamefont{Altland}} \bibnamefont{and}
  \bibinfo{author}{\bibfnamefont{T.}~\bibnamefont{Micklitz}},
  \bibinfo{journal}{Phys. Rev. Lett.} \textbf{\bibinfo{volume}{118}},
  \bibinfo{pages}{127202} (\bibinfo{year}{2017}).

\bibitem[{\citenamefont{Kondov et~al.}(2015)\citenamefont{Kondov, McGehee, Xu,
  and DeMarco}}]{Kondov:2015aa}
\bibinfo{author}{\bibfnamefont{S.~S.} \bibnamefont{Kondov}},
  \bibinfo{author}{\bibfnamefont{W.~R.} \bibnamefont{McGehee}},
  \bibinfo{author}{\bibfnamefont{W.}~\bibnamefont{Xu}}, \bibnamefont{and}
  \bibinfo{author}{\bibfnamefont{B.}~\bibnamefont{DeMarco}},
  \bibinfo{journal}{Phys. Rev. Lett.} \textbf{\bibinfo{volume}{114}},
  \bibinfo{pages}{083002} (\bibinfo{year}{2015}).

\bibitem[{\citenamefont{Schreiber et~al.}(2015)\citenamefont{Schreiber,
  Hodgman, Bordia, L{\"u}schen, Fischer, Vosk, Altman, Schneider, and
  Bloch}}]{Schreiber:2015aa}
\bibinfo{author}{\bibfnamefont{M.}~\bibnamefont{Schreiber}},
  \bibinfo{author}{\bibfnamefont{S.~S.} \bibnamefont{Hodgman}},
  \bibinfo{author}{\bibfnamefont{P.}~\bibnamefont{Bordia}},
  \bibinfo{author}{\bibfnamefont{H.~P.} \bibnamefont{L{\"u}schen}},
  \bibinfo{author}{\bibfnamefont{M.~H.} \bibnamefont{Fischer}},
  \bibinfo{author}{\bibfnamefont{R.}~\bibnamefont{Vosk}},
  \bibinfo{author}{\bibfnamefont{E.}~\bibnamefont{Altman}},
  \bibinfo{author}{\bibfnamefont{U.}~\bibnamefont{Schneider}},
  \bibnamefont{and} \bibinfo{author}{\bibfnamefont{I.}~\bibnamefont{Bloch}},
  \bibinfo{journal}{Science} \textbf{\bibinfo{volume}{349}},
  \bibinfo{pages}{842} (\bibinfo{year}{2015}).

\bibitem[{\citenamefont{Choi et~al.}(2016)\citenamefont{Choi, Hild, Zeiher,
  Schau{\ss}, Rubio-Abadal, Yefsah, Khemani, Huse, Bloch, and
  Gross}}]{Choi:2016aa}
\bibinfo{author}{\bibfnamefont{J.-y.} \bibnamefont{Choi}},
  \bibinfo{author}{\bibfnamefont{S.}~\bibnamefont{Hild}},
  \bibinfo{author}{\bibfnamefont{J.}~\bibnamefont{Zeiher}},
  \bibinfo{author}{\bibfnamefont{P.}~\bibnamefont{Schau{\ss}}},
  \bibinfo{author}{\bibfnamefont{A.}~\bibnamefont{Rubio-Abadal}},
  \bibinfo{author}{\bibfnamefont{T.}~\bibnamefont{Yefsah}},
  \bibinfo{author}{\bibfnamefont{V.}~\bibnamefont{Khemani}},
  \bibinfo{author}{\bibfnamefont{D.~A.} \bibnamefont{Huse}},
  \bibinfo{author}{\bibfnamefont{I.}~\bibnamefont{Bloch}}, \bibnamefont{and}
  \bibinfo{author}{\bibfnamefont{C.}~\bibnamefont{Gross}},
  \bibinfo{journal}{Science} \textbf{\bibinfo{volume}{352}},
  \bibinfo{pages}{1547} (\bibinfo{year}{2016}).

\bibitem[{\citenamefont{{Bordia} et~al.}(2017)\citenamefont{{Bordia},
  {L{\"u}schen}, {Scherg}, {Gopalakrishnan}, {Knap}, {Schneider}, and
  {Bloch}}}]{Bordia:2017aa}
\bibinfo{author}{\bibfnamefont{P.}~\bibnamefont{{Bordia}}},
  \bibinfo{author}{\bibfnamefont{H.}~\bibnamefont{{L{\"u}schen}}},
  \bibinfo{author}{\bibfnamefont{S.}~\bibnamefont{{Scherg}}},
  \bibinfo{author}{\bibfnamefont{S.}~\bibnamefont{{Gopalakrishnan}}},
  \bibinfo{author}{\bibfnamefont{M.}~\bibnamefont{{Knap}}},
  \bibinfo{author}{\bibfnamefont{U.}~\bibnamefont{{Schneider}}},
  \bibnamefont{and} \bibinfo{author}{\bibfnamefont{I.}~\bibnamefont{{Bloch}}},
  \bibinfo{journal}{ArXiv e-prints}  (\bibinfo{year}{2017}),
  \eprint{1704.03063}.

\bibitem[{\citenamefont{Smith et~al.}(2016)\citenamefont{Smith, Lee, Richerme,
  Neyenhuis, Hess, Hauke, Heyl, Huse, and Monroe}}]{Smith:2016aa}
\bibinfo{author}{\bibfnamefont{J.}~\bibnamefont{Smith}},
  \bibinfo{author}{\bibfnamefont{A.}~\bibnamefont{Lee}},
  \bibinfo{author}{\bibfnamefont{P.}~\bibnamefont{Richerme}},
  \bibinfo{author}{\bibfnamefont{B.}~\bibnamefont{Neyenhuis}},
  \bibinfo{author}{\bibfnamefont{P.~W.} \bibnamefont{Hess}},
  \bibinfo{author}{\bibfnamefont{P.}~\bibnamefont{Hauke}},
  \bibinfo{author}{\bibfnamefont{M.}~\bibnamefont{Heyl}},
  \bibinfo{author}{\bibfnamefont{D.~A.} \bibnamefont{Huse}}, \bibnamefont{and}
  \bibinfo{author}{\bibfnamefont{C.}~\bibnamefont{Monroe}},
  \bibinfo{journal}{Nat Phys} \textbf{\bibinfo{volume}{12}},
  \bibinfo{pages}{907} (\bibinfo{year}{2016}).

\bibitem[{\citenamefont{{Kucsko} et~al.}(2016)\citenamefont{{Kucsko}, {Choi},
  {Choi}, {Maurer}, {Sumiya}, {Onoda}, {Isoya}, {Jelezko}, {Demler}, {Yao}
  et~al.}}]{Kucsko:2016aa}
\bibinfo{author}{\bibfnamefont{G.}~\bibnamefont{{Kucsko}}},
  \bibinfo{author}{\bibfnamefont{S.}~\bibnamefont{{Choi}}},
  \bibinfo{author}{\bibfnamefont{J.}~\bibnamefont{{Choi}}},
  \bibinfo{author}{\bibfnamefont{P.~C.} \bibnamefont{{Maurer}}},
  \bibinfo{author}{\bibfnamefont{H.}~\bibnamefont{{Sumiya}}},
  \bibinfo{author}{\bibfnamefont{S.}~\bibnamefont{{Onoda}}},
  \bibinfo{author}{\bibfnamefont{J.}~\bibnamefont{{Isoya}}},
  \bibinfo{author}{\bibfnamefont{F.}~\bibnamefont{{Jelezko}}},
  \bibinfo{author}{\bibfnamefont{E.}~\bibnamefont{{Demler}}},
  \bibinfo{author}{\bibfnamefont{N.~Y.} \bibnamefont{{Yao}}},
  \bibnamefont{et~al.}, \bibinfo{journal}{ArXiv e-prints}
  (\bibinfo{year}{2016}), \eprint{1609.08216}.

\bibitem[{\citenamefont{Chandran et~al.}(2016)\citenamefont{Chandran, Pal,
  Laumann, and Scardicchio}}]{Chandran:2016ac}
\bibinfo{author}{\bibfnamefont{A.}~\bibnamefont{Chandran}},
  \bibinfo{author}{\bibfnamefont{A.}~\bibnamefont{Pal}},
  \bibinfo{author}{\bibfnamefont{C.~R.} \bibnamefont{Laumann}},
  \bibnamefont{and}
  \bibinfo{author}{\bibfnamefont{A.}~\bibnamefont{Scardicchio}},
  \bibinfo{journal}{Phys. Rev. B} \textbf{\bibinfo{volume}{94}},
  \bibinfo{pages}{144203} (\bibinfo{year}{2016}).

\bibitem[{\citenamefont{De~Roeck and Huveneers}(2017)}]{De-Roeck:2017aa}
\bibinfo{author}{\bibfnamefont{W.}~\bibnamefont{De~Roeck}} \bibnamefont{and}
  \bibinfo{author}{\bibfnamefont{F.}~\bibnamefont{Huveneers}},
  \bibinfo{journal}{Physical Review B} \textbf{\bibinfo{volume}{95}}
  (\bibinfo{year}{2017}), ISSN \bibinfo{issn}{2469-9969}.

\bibitem[{\citenamefont{{Luitz} et~al.}(2017)\citenamefont{{Luitz},
  {Huveneers}, and {de Roeck}}}]{Luitz:2017aa}
\bibinfo{author}{\bibfnamefont{D.~J.} \bibnamefont{{Luitz}}},
  \bibinfo{author}{\bibfnamefont{F.}~\bibnamefont{{Huveneers}}},
  \bibnamefont{and} \bibinfo{author}{\bibfnamefont{W.}~\bibnamefont{{de
  Roeck}}}, \bibinfo{journal}{ArXiv e-prints}  (\bibinfo{year}{2017}),
  \eprint{1705.10807}.

\bibitem[{\citenamefont{Cugliandolo et~al.}(2001)\citenamefont{Cugliandolo,
  Grempel, and da~Silva~Santos}}]{Cugliandolo:2001aa}
\bibinfo{author}{\bibfnamefont{L.~F.} \bibnamefont{Cugliandolo}},
  \bibinfo{author}{\bibfnamefont{D.~R.} \bibnamefont{Grempel}},
  \bibnamefont{and} \bibinfo{author}{\bibfnamefont{C.~A.}
  \bibnamefont{da~Silva~Santos}}, \bibinfo{journal}{Phys. Rev. B}
  \textbf{\bibinfo{volume}{64}}, \bibinfo{pages}{014403}
  (\bibinfo{year}{2001}).

\bibitem[{\citenamefont{Laumann et~al.}(2014)\citenamefont{Laumann, Pal, and
  Scardicchio}}]{Laumann:2014aa}
\bibinfo{author}{\bibfnamefont{C.~R.} \bibnamefont{Laumann}},
  \bibinfo{author}{\bibfnamefont{A.}~\bibnamefont{Pal}}, \bibnamefont{and}
  \bibinfo{author}{\bibfnamefont{A.}~\bibnamefont{Scardicchio}},
  \bibinfo{journal}{Phys. Rev. Lett.} \textbf{\bibinfo{volume}{113}},
  \bibinfo{pages}{200405} (\bibinfo{year}{2014}).

\bibitem[{\citenamefont{{Burin}}(2016)}]{Burin:2016aa}
\bibinfo{author}{\bibfnamefont{A.~L.} \bibnamefont{{Burin}}},
  \bibinfo{journal}{ArXiv e-prints}  (\bibinfo{year}{2016}),
  \eprint{1610.00811}.

\bibitem[{\citenamefont{Baldwin et~al.}(2017)\citenamefont{Baldwin, Laumann,
  Pal, and Scardicchio}}]{Baldwin:2017aa}
\bibinfo{author}{\bibfnamefont{C.}~\bibnamefont{Baldwin}},
  \bibinfo{author}{\bibfnamefont{C.}~\bibnamefont{Laumann}},
  \bibinfo{author}{\bibfnamefont{A.}~\bibnamefont{Pal}}, \bibnamefont{and}
  \bibinfo{author}{\bibfnamefont{A.}~\bibnamefont{Scardicchio}},
  \bibinfo{journal}{Phys. Rev. Lett.} \textbf{\bibinfo{volume}{118}}
  (\bibinfo{year}{2017}), ISSN \bibinfo{issn}{1079-7114}.

\bibitem[{\citenamefont{M\'ezard et~al.}(1987)\citenamefont{M\'ezard, Parisi,
  and Virasoro}}]{Mezard:1987aa}
\bibinfo{author}{\bibfnamefont{M.}~\bibnamefont{M\'ezard}},
  \bibinfo{author}{\bibfnamefont{G.}~\bibnamefont{Parisi}}, \bibnamefont{and}
  \bibinfo{author}{\bibfnamefont{M.}~\bibnamefont{Virasoro}},
  \emph{\bibinfo{title}{Spin Glass Theory and Beyond}}, Lecture Notes in
  Physics Series (\bibinfo{publisher}{World Scientific Publishing Company,
  Incorporated}, \bibinfo{year}{1987}), ISBN \bibinfo{isbn}{9789971501150}.

\bibitem[{\citenamefont{Yu et~al.}(2016)\citenamefont{Yu, Luitz, and
  Clark}}]{Yu:2016aa}
\bibinfo{author}{\bibfnamefont{X.}~\bibnamefont{Yu}},
  \bibinfo{author}{\bibfnamefont{D.~J.} \bibnamefont{Luitz}}, \bibnamefont{and}
  \bibinfo{author}{\bibfnamefont{B.~K.} \bibnamefont{Clark}},
  \bibinfo{journal}{Phys. Rev. B} \textbf{\bibinfo{volume}{94}},
  \bibinfo{pages}{184202} (\bibinfo{year}{2016}).

\bibitem[{\citenamefont{Khemani et~al.}(2017)\citenamefont{Khemani, Lim, Sheng,
  and Huse}}]{Khemani:2016aa}
\bibinfo{author}{\bibfnamefont{V.}~\bibnamefont{Khemani}},
  \bibinfo{author}{\bibfnamefont{S.~P.} \bibnamefont{Lim}},
  \bibinfo{author}{\bibfnamefont{D.~N.} \bibnamefont{Sheng}}, \bibnamefont{and}
  \bibinfo{author}{\bibfnamefont{D.~A.} \bibnamefont{Huse}},
  \bibinfo{journal}{Phys. Rev. X} \textbf{\bibinfo{volume}{7}},
  \bibinfo{pages}{021013} (\bibinfo{year}{2017}).

\bibitem[{\citenamefont{Altshuler et~al.}(1997)\citenamefont{Altshuler, Gefen,
  Kamenev, and Levitov}}]{Altshuler:1997aa}
\bibinfo{author}{\bibfnamefont{B.~L.} \bibnamefont{Altshuler}},
  \bibinfo{author}{\bibfnamefont{Y.}~\bibnamefont{Gefen}},
  \bibinfo{author}{\bibfnamefont{A.}~\bibnamefont{Kamenev}}, \bibnamefont{and}
  \bibinfo{author}{\bibfnamefont{L.~S.} \bibnamefont{Levitov}},
  \bibinfo{journal}{Phys. Rev. Lett.} \textbf{\bibinfo{volume}{78}},
  \bibinfo{pages}{2803} (\bibinfo{year}{1997}).

\bibitem[{\citenamefont{Atas et~al.}(2013)\citenamefont{Atas, Bogomolny,
  Giraud, and Roux}}]{Atas:2013aa}
\bibinfo{author}{\bibfnamefont{Y.~Y.} \bibnamefont{Atas}},
  \bibinfo{author}{\bibfnamefont{E.}~\bibnamefont{Bogomolny}},
  \bibinfo{author}{\bibfnamefont{O.}~\bibnamefont{Giraud}}, \bibnamefont{and}
  \bibinfo{author}{\bibfnamefont{G.}~\bibnamefont{Roux}},
  \bibinfo{journal}{Phys. Rev. Lett.} \textbf{\bibinfo{volume}{110}},
  \bibinfo{pages}{084101} (\bibinfo{year}{2013}).

\bibitem[{\citenamefont{Chayes et~al.}(1986)\citenamefont{Chayes, Chayes,
  Fisher, and Spencer}}]{Chayes:1986kq}
\bibinfo{author}{\bibfnamefont{J.~T.} \bibnamefont{Chayes}},
  \bibinfo{author}{\bibfnamefont{L.}~\bibnamefont{Chayes}},
  \bibinfo{author}{\bibfnamefont{D.~S.} \bibnamefont{Fisher}},
  \bibnamefont{and} \bibinfo{author}{\bibfnamefont{T.}~\bibnamefont{Spencer}},
  \bibinfo{journal}{Phys. Rev. Lett.} \textbf{\bibinfo{volume}{57}},
  \bibinfo{pages}{2999} (\bibinfo{year}{1986}).

\bibitem[{\citenamefont{{Chandran} et~al.}(2015)\citenamefont{{Chandran},
  {Laumann}, and {Oganesyan}}}]{Chandran:2015aa}
\bibinfo{author}{\bibfnamefont{A.}~\bibnamefont{{Chandran}}},
  \bibinfo{author}{\bibfnamefont{C.~R.} \bibnamefont{{Laumann}}},
  \bibnamefont{and}
  \bibinfo{author}{\bibfnamefont{V.}~\bibnamefont{{Oganesyan}}},
  \bibinfo{journal}{ArXiv e-prints}  (\bibinfo{year}{2015}),
  \eprint{1509.04285}.

\bibitem[{\citenamefont{Vosk et~al.}(2015)\citenamefont{Vosk, Huse, and
  Altman}}]{Vosk:2015aa}
\bibinfo{author}{\bibfnamefont{R.}~\bibnamefont{Vosk}},
  \bibinfo{author}{\bibfnamefont{D.~A.} \bibnamefont{Huse}}, \bibnamefont{and}
  \bibinfo{author}{\bibfnamefont{E.}~\bibnamefont{Altman}},
  \bibinfo{journal}{Phys. Rev. X} \textbf{\bibinfo{volume}{5}},
  \bibinfo{pages}{031032} (\bibinfo{year}{2015}).

\bibitem[{\citenamefont{Potter et~al.}(2015)\citenamefont{Potter, Vasseur, and
  Parameswaran}}]{Potter:2015ab}
\bibinfo{author}{\bibfnamefont{A.~C.} \bibnamefont{Potter}},
  \bibinfo{author}{\bibfnamefont{R.}~\bibnamefont{Vasseur}}, \bibnamefont{and}
  \bibinfo{author}{\bibfnamefont{S.~A.} \bibnamefont{Parameswaran}},
  \bibinfo{journal}{Phys. Rev. X} \textbf{\bibinfo{volume}{5}},
  \bibinfo{pages}{031033} (\bibinfo{year}{2015}).

\bibitem[{\citenamefont{Borgs et~al.}(2001)\citenamefont{Borgs, Chayes, Kesten,
  and Spencer}}]{Borgs:2001fw}
\bibinfo{author}{\bibfnamefont{C.}~\bibnamefont{Borgs}},
  \bibinfo{author}{\bibfnamefont{J.~T.} \bibnamefont{Chayes}},
  \bibinfo{author}{\bibfnamefont{H.}~\bibnamefont{Kesten}}, \bibnamefont{and}
  \bibinfo{author}{\bibfnamefont{J.}~\bibnamefont{Spencer}},
  \bibinfo{journal}{Communications in Mathematical Physics}
  \textbf{\bibinfo{volume}{224}}, \bibinfo{pages}{153} (\bibinfo{year}{2001}).

\bibitem[{\citenamefont{Heydenreich and van~der
  Hofstad}(2015)}]{Heydenreich:2015aa}
\bibinfo{author}{\bibfnamefont{M.}~\bibnamefont{Heydenreich}} \bibnamefont{and}
  \bibinfo{author}{\bibfnamefont{R.}~\bibnamefont{van~der Hofstad}},
  \bibinfo{journal}{Lecture notes for the CRM-PIMS Summer School in
  Probability}  (\bibinfo{year}{2015}).

\bibitem[{\citenamefont{Bollob{\'a}s}(2001)}]{Bollobas:2001sj}
\bibinfo{author}{\bibfnamefont{B.}~\bibnamefont{Bollob{\'a}s}},
  \emph{\bibinfo{title}{Random Graphs}} (\bibinfo{publisher}{Cambridge
  University Press}, \bibinfo{address}{Cambridge, UK}, \bibinfo{year}{2001}),
  ISBN \bibinfo{isbn}{0521797225}.

\bibitem[{\citenamefont{Abou-Chacra et~al.}(1973)\citenamefont{Abou-Chacra,
  Thouless, and Anderson}}]{Abou-Chacra:1973qf}
\bibinfo{author}{\bibfnamefont{R.}~\bibnamefont{Abou-Chacra}},
  \bibinfo{author}{\bibfnamefont{D.~J.} \bibnamefont{Thouless}},
  \bibnamefont{and} \bibinfo{author}{\bibfnamefont{P.~W.}
  \bibnamefont{Anderson}}, \bibinfo{journal}{Journal of Physics C: Solid State
  Physics} \textbf{\bibinfo{volume}{6}}, \bibinfo{pages}{1734}
  (\bibinfo{year}{1973}).

\bibitem[{\citenamefont{Biroli et~al.}(2012)\citenamefont{Biroli,
  Ribeiro-Teixeira, and Tarzia}}]{Biroli:2012vk}
\bibinfo{author}{\bibfnamefont{G.}~\bibnamefont{Biroli}},
  \bibinfo{author}{\bibfnamefont{A.~C.} \bibnamefont{Ribeiro-Teixeira}},
  \bibnamefont{and} \bibinfo{author}{\bibfnamefont{M.}~\bibnamefont{Tarzia}},
  \bibinfo{journal}{arXiv.org}  (\bibinfo{year}{2012}), \eprint{1211.7334v2}.

\bibitem[{\citenamefont{De~Luca et~al.}(2014)\citenamefont{De~Luca, Altshuler,
  Kravtsov, and Scardicchio}}]{DeLuca:2014ch}
\bibinfo{author}{\bibfnamefont{A.}~\bibnamefont{De~Luca}},
  \bibinfo{author}{\bibfnamefont{B.~L.} \bibnamefont{Altshuler}},
  \bibinfo{author}{\bibfnamefont{V.~E.} \bibnamefont{Kravtsov}},
  \bibnamefont{and}
  \bibinfo{author}{\bibfnamefont{A.}~\bibnamefont{Scardicchio}},
  \bibinfo{journal}{Phys. Rev. Lett.} \textbf{\bibinfo{volume}{113}},
  \bibinfo{pages}{046806} (\bibinfo{year}{2014}).

\bibitem[{\citenamefont{Altshuler et~al.}(2016)\citenamefont{Altshuler, Cuevas,
  Ioffe, and Kravtsov}}]{Altshuler:2016aa}
\bibinfo{author}{\bibfnamefont{B.~L.} \bibnamefont{Altshuler}},
  \bibinfo{author}{\bibfnamefont{E.}~\bibnamefont{Cuevas}},
  \bibinfo{author}{\bibfnamefont{L.~B.} \bibnamefont{Ioffe}}, \bibnamefont{and}
  \bibinfo{author}{\bibfnamefont{V.~E.} \bibnamefont{Kravtsov}},
  \bibinfo{journal}{Phys. Rev. Lett.} \textbf{\bibinfo{volume}{117}},
  \bibinfo{pages}{156601} (\bibinfo{year}{2016}).

\bibitem[{\citenamefont{Fisher}(1995)}]{Fisher:1995oq}
\bibinfo{author}{\bibfnamefont{D.~S.} \bibnamefont{Fisher}},
  \bibinfo{journal}{Phys. Rev. B} \textbf{\bibinfo{volume}{51}},
  \bibinfo{pages}{6411} (\bibinfo{year}{1995}).

\bibitem[{\citenamefont{P\'azm\'andi et~al.}(1997)\citenamefont{P\'azm\'andi,
  Scalettar, and Zim\'anyi}}]{Pazmandi:1997aa}
\bibinfo{author}{\bibfnamefont{F.}~\bibnamefont{P\'azm\'andi}},
  \bibinfo{author}{\bibfnamefont{R.~T.} \bibnamefont{Scalettar}},
  \bibnamefont{and} \bibinfo{author}{\bibfnamefont{G.~T.}
  \bibnamefont{Zim\'anyi}}, \bibinfo{journal}{Phys. Rev. Lett.}
  \textbf{\bibinfo{volume}{79}}, \bibinfo{pages}{5130} (\bibinfo{year}{1997}).

\bibitem[{\citenamefont{Deutsch}(1991)}]{Deutsch:1991ss}
\bibinfo{author}{\bibfnamefont{J.~M.} \bibnamefont{Deutsch}},
  \bibinfo{journal}{Phys. Rev. A} \textbf{\bibinfo{volume}{43}},
  \bibinfo{pages}{2046} (\bibinfo{year}{1991}).

\bibitem[{\citenamefont{Srednicki}(1994)}]{Srednicki:1994dw}
\bibinfo{author}{\bibfnamefont{M.}~\bibnamefont{Srednicki}},
  \bibinfo{journal}{Phys. Rev. E} \textbf{\bibinfo{volume}{50}},
  \bibinfo{pages}{888} (\bibinfo{year}{1994}).

\bibitem[{\citenamefont{Rigol et~al.}(2008)\citenamefont{Rigol, Dunjko, and
  Olshanii}}]{Rigol:2008bh}
\bibinfo{author}{\bibfnamefont{M.}~\bibnamefont{Rigol}},
  \bibinfo{author}{\bibfnamefont{V.}~\bibnamefont{Dunjko}}, \bibnamefont{and}
  \bibinfo{author}{\bibfnamefont{M.}~\bibnamefont{Olshanii}},
  \bibinfo{journal}{Nature} \textbf{\bibinfo{volume}{452}},
  \bibinfo{pages}{854} (\bibinfo{year}{2008}).

\bibitem[{\citenamefont{D'Alessio et~al.}(2016)\citenamefont{D'Alessio, Kafri,
  Polkovnikov, and Rigol}}]{DAlessio:2016aa}
\bibinfo{author}{\bibfnamefont{L.}~\bibnamefont{D'Alessio}},
  \bibinfo{author}{\bibfnamefont{Y.}~\bibnamefont{Kafri}},
  \bibinfo{author}{\bibfnamefont{A.}~\bibnamefont{Polkovnikov}},
  \bibnamefont{and} \bibinfo{author}{\bibfnamefont{M.}~\bibnamefont{Rigol}},
  \bibinfo{journal}{Advances in Physics} \textbf{\bibinfo{volume}{65}},
  \bibinfo{pages}{239} (\bibinfo{year}{2016}), ISSN \bibinfo{issn}{1460-6976}.

\bibitem[{\citenamefont{Borgonovi et~al.}(2016)\citenamefont{Borgonovi,
  Izrailev, Santos, and Zelevinsky}}]{Borgonovi:2016aa}
\bibinfo{author}{\bibfnamefont{F.}~\bibnamefont{Borgonovi}},
  \bibinfo{author}{\bibfnamefont{F.~M.} \bibnamefont{Izrailev}},
  \bibinfo{author}{\bibfnamefont{L.~F.} \bibnamefont{Santos}},
  \bibnamefont{and} \bibinfo{author}{\bibfnamefont{V.~G.}
  \bibnamefont{Zelevinsky}}, \bibinfo{journal}{Physics Reports}
  \textbf{\bibinfo{volume}{626}}, \bibinfo{pages}{1} (\bibinfo{year}{2016}).

\end{thebibliography}

\end{document}